\documentclass[11pt,a4paper]{article}
\usepackage[latin1]{inputenc}
\usepackage{pdflscape}

\usepackage{graphicx,booktabs}
\usepackage{multicol,wrapfig}
\usepackage{amsfonts}
\usepackage{amsmath}
\usepackage{amssymb}
\usepackage{theorem}
\usepackage{upgreek}
\usepackage{xcolor}
\usepackage{float}

\theoremstyle{change}
\theorembodyfont{\slshape}
\newtheorem{satz}{Theorem}[section]

\newtheorem{prop}[satz]{Proposition}

\theorembodyfont{\upshape\small}

\newtheorem{bem}[satz]{Remark}

\theorembodyfont{\ttfamily}

\newcommand{\ba}{\begin{equation}}
\newcommand{\ea}{\end{equation}}

\usepackage{dsfont}

\newcommand{\0}{\mbox{\boldmath $0$}}
\newcommand{\1}{\mbox{\boldmath $1$}}
\newcommand{\fa}{\mbox{\boldmath $a$}}

\newcommand{\fh}{\mbox{\boldmath $h$}}
\newcommand{\fhi}{\mbox{\boldmath\scriptsize $h$}}

\newcommand{\fp}{\mbox{\boldmath $p$}}

\newcommand{\ft}{\mbox{\boldmath $t$}}
\newcommand{\fti}{\mbox{\boldmath\scriptsize $t$}}

\newcommand{\fX}{\mbox{\boldmath $X$}}
\newcommand{\fx}{\mbox{\boldmath $x$}}

\newcommand{\fmu}{\mbox{\boldmath $\mu$}}

\newcommand{\mC}{\mbox{\textup{\textbf{C}}}}

\newcommand{\ftau}{\mbox{\boldmath $\tau$}}
\newcommand{\ftaui}{\mbox{\boldmath\scriptsize $\tau$}}
\newcommand{\fSigma}{\mbox{\boldmath $\Sigma$}}

\newcommand{\norm}{\textup{N}}

\newcommand{\bbn}{\mathbb{N}}
\newcommand{\bbr}{\mathbb{R}}
\newcommand{\bbz}{\mathbb{Z}}

\newcommand{\cat}{m}

\newcommand{\iid}{i.\,i.\,d.}
\newcommand{\ie}{i.\,e., }
\newcommand{\eg}{e.\,g., }

\usepackage{natbib}
\usepackage{hyperref}
\begin{document}
%\pagestyle{headings}

%\pagenumbering{roman}

%\pretolerance=10000000
%Verhindert Silbentrennung

\parindent 0cm
%Verhindert Einr?ken der Abs?ze

\title{Nonparametric Testing of Spatial Dependence in 2D and 3D Random Fields}

\author{
Christian H.\ Wei\ss{}\thanks{Department of Mathematics and Statistics, Helmut Schmidt University, 22043 Hamburg, Germany}\ \thanks{Corresponding author. E-Mail: \href{mailto:weissc@hsu-hh.de}{\nolinkurl{weissc@hsu-hh.de}}. ORCID: \href{https://orcid.org/0000-0001-8739-6631}{\nolinkurl{0000-0001-8739-6631}}.}
\and 
Philipp Ad\"ammer\thanks{Institute of Data Science, University of Greifswald, 17489 Greifswald, Germany.}\ \thanks{E-Mail: \href{mailto:philipp.adaemmer@uni-greifswald.de}{\nolinkurl{philipp.adaemmer@uni-greifswald.de}}. ORCID: \href{https://orcid.org/0000-0003-3770-0097}{\nolinkurl{0000-0003-3770-0097}}.}
}

\maketitle

\begin{abstract}

We propose a flexible and robust nonparametric framework for testing spatial dependence in two- and three-dimensional random fields. Our approach involves converting spatial data into one-dimensional time series using space-filling Hilbert curves. We then apply ordinal pattern-based tests for serial dependence to this series. Because Hilbert curves preserve spatial locality, spatial dependence in the original field manifests as serial dependence in the transformed sequence. The approach is easy to implement, accommodates arbitrary grid sizes through generalized Hilbert (``gilbert'') curves, and naturally extends beyond three dimensions. This provides a practical and general alternative to existing methods based on spatial ordinal patterns, which are typically limited to two-dimensional settings.

\end{abstract}

\section{Introduction}
\label{Introduction}
In their seminal paper, \citet{bandt02} introduced ordinal patterns (OPs) as a rank-based approach for time series analysis. Since that time, OP-methods found various applications in practice, and several refinements and extensions have been proposed in the literature, see \citet{bandt19,bandt23} for a survey about the state-of-art and recent references. Among others, OPs were used to develop nonparametric (and robust) tests of serial dependence, see \citet{weiss22} for a comprehensive discussion. 
In the present research, however, we do not focus on time-series data but rather on spatial data observed on a regular grid, which are generated by a random field. Our objective is to develop and investigate a ``flexible'' and robust nonparametric approach for testing spatial dependence. The term flexible here refers to the dimensionality of the spatial data. While most of the existing literature concentrates on the two-dimensional (2D) case (see details below), the approach proposed in this study is applicable to higher-dimensional settings as well. In particular, we examine the three-dimensional (3D) case in depth, extending beyond the 2D focus of previous studies.

\smallskip
For the 2D case, there exists a tailor-made OP-based solution for spatial data analysis, namely the concept of spatial OPs (SOPs). SOPs have been proposed by \citet{ribeiro12} and were further applied by \citet{zunino16,sigaki18}. A comprehensive analysis and further developments of SOPs are by \citet{bandtw23}, who introduced, among others, the concept of the ``type'' of a SOP. Finally, \citet{WeissKim2024,WeissKim2025} developed and investigated nonparametric tests for spatial dependence in 2D random fields, which rely on an asymptotic implementation of various SOP-based statistics. However, the necessary derivations are rather sophisticated such that an extensions to higher dimensions does not appear to be feasible. In fact, also the extension of SOPs themselves to higher dimensions is not straightforward due to the additional complexity caused by increasing dimensionality.

\smallskip
For these reasons, our proposal takes a different approach that can easily be extended to higher dimensions without the need for further asymptotic derivations. We suggest using Hilbert curves to transform spatial data into a time series and applying the tests of \citet{weiss22} to the resulting time series. Hilbert curves are a particular type of space-filling curve; the 2D version was first described by \citet{hilbert91}. Since Hilbert curves are known to preserve locality (``clustering property''), see \citet{jagadish90,moon01}, we conjecture that spatial dependence in the original data translates into serial dependence in the transformed data. The latter can then be uncovered by tests based on ordinary OPs for time series.
Furthermore, Hilbert curves are defined not only for 2D, but also for 3D \citep{sagan93} and even higher grid data \citep{jagadish90,moon01}. Thus, our proposed approach is universally applicable, irrespective of the dimension of the grid. In fact, the idea to use Hilbert curves together with OPs is not completely new, because \citet{chagas21,bariviera25} successfully applied Hilbert curves to 2D image data for texture analysis. But here, we go a step further by testing for spatial dependence in regular grid data, and by also considering higher dimensions than just~2D. Furthermore, the aforementioned papers focused on quadratic datasets of size $2^r\times2^r$, because Hilbert curves were originally developed only for quadratic or cubic data where the length is a power of two. Also standard software implementations such as the command \href{https://reference.wolfram.com/language/ref/HilbertCurve.html}{\nolinkurl{HilbertCurve}} in Wolfram Mathematica are limited to this case (but are flexible in terms of dimensionality). In the present research, we make use of recent proposals for constructing Hilbert curves for arbitrary-sized rectangles or cuboids, so-called generalized Hilbert (``gilbert'') curves, such as developed by \citet{rong21} or as offered by the Github project \href{https://github.com/jakubcerveny/gilbert}{\nolinkurl{gilbert}} by Jakub \v{C}erven\'y, also see the \texttt{Julia} package \href{https://github.com/CliMA/GilbertCurves.jl}{\nolinkurl{GilbertCurves}} (both accessed on May 26, 2025).

\smallskip
In what follows, we propose nonparametric Hilbert-OP tests for spatial dependence in 2D and 3D grid data, which could be straightforwardly extended to even higher dimensions. In particular, we analyze their power performance for various kinds of data-generating process (DGP), namely different 2D and 3D stationary random fields. 
The remainder of the paper is structured as follows. Section~\ref{Hilbert-Curve Approach for Spatial Dependence Testing} introduces Hilbert curves and demonstrates how 2D and 3D grid data can be transformed into a univariate time series. Section~\ref{Performance Analyses} presents an extensive simulation study that investigates the power of various statistical tests under different data-generating processes (DGPs). Section~\ref{Data Applications} applies the proposed Hilbert curve approach to both 2D and 3D data, specifically to art paintings (3D) and barley yield (2D). Finally, Section~\ref{Conclusions} concludes the paper and outlines directions for future research.

\numberwithin{satz}{subsection}

\section{Hilbert-Curve Approach for Spatial Dependence Testing}
\label{Hilbert-Curve Approach for Spatial Dependence Testing}

\subsection{Summary of Relevant Previous Results}
\label{Summary of Relevant Previous Results}
Let $(X_t)_{t\in\bbz=\{\ldots,-1,0,1,\ldots\}}$ be a real-valued and continuously distributed process, and let $\fX_t=(X_t,X_{t+d},\ldots,X_{t+(\cat-1)d})$ be the $t$th segment of length $\cat\in\bbn=\{1,2,\ldots\}$, $\cat\geq 2$, and with delay $d\in\bbn$. Then, $\fX_t$ can be mapped onto a permutation $\uppi_t\in S_\cat = \{\uppi^{[1]},\ldots,\uppi^{[\cat!]}\}$ from the symmetric group~$S_\cat$ of order~$\cat$ by defining~$\uppi_t$ to be the OP of~$\fX_t$. Here, we say that $\uppi=(r_1,\ldots,r_\cat)\in S_\cat$ is the OP of $\fx=(x_1,\ldots,x_\cat)\in\bbr^\cat$ if it satisfies
\ba
\label{ordpatternrank}
r_k<r_l\qquad\Leftrightarrow\qquad
x_{k}< x_{l} \quad\text{or}\quad
(x_k=x_l \text{ and } k<l)
\ea
for all $k,l\in \{1,\ldots,\cat\}$, \ie the integers within~$\uppi$ express the ranks of the entries of~$\fx$. The case ``$x_k=x_l$'' in \eqref{ordpatternrank} handles the possible occurrence of ties within~$\fx$, although our requirement for a continuously distributed process $(X_t)$ implies that ties happen with probability zero. Note that several (equivalent) definitions of ordinal patterns (OPs) exist; see \citet{berger19} for an overview. However, we adopt the rank-based representation in \eqref{ordpatternrank} for its ease of interpretation. Also note that since OPs rely solely on ranks, OP-based statistics are inherently robust to outliers.

\begin{bem}
\label{bemTies}
If $(X_t)$ would be a discretely distributed (and at least ordinal) process, then the probability for ties would be truly positive, and corresponding time-series data may exhibit frequent ties. If one wants to explicitly consider such ties, then one can use so-called generalized OPs (GOPs) as first proposed by \citet{bian12,unakafova13}. However, the corresponding hypothesis tests for serial dependence, see \citet{WeissSchnurr2024} for details, are quite different from those presented below. Therefore, for the sake of conciseness, we shall use a randomization approach instead to apply the subsequent OP-tests directly to time series exhibiting frequent ties. For example, if $(X_t)$ is integer-valued, then we first add standard-uniform noise to the data before computing the ordinary OPs and related statistics, see \citet{WeissSchnurr2024} for performance comparisons. The combination of GOPs with the Hilbert-curve approach proposed in Section~\ref{The Novel Hilbert-Curve Approach} is recommended as an interesting task for future research.
\end{bem}
The importance of the OPs \eqref{ordpatternrank} for dependence testing arises from the fact that under the null hypothesis of serial independence, \ie if $(X_t)$ is independent and identically distributed (\iid), the marginal distribution of the OP~series $(\uppi_t)$ is simply the discrete-uniform distribution with $P(\uppi_t=\uppi) = 1/\cat!$ for all $\uppi\in S_\cat$, irrespective of the distribution of~$X_t$. So OPs lead to nonparametric (distribution-free) approaches under the \iid-null. In the presence of serial dependence, by contrast, we typically observe deviations from uniformity, see \citet{bandt07,sousa22,silbernagel25} for details. Hence, testing for serial dependence can be implemented by checking the OPs' marginal distribution for deviations from uniformity. This is typically done by computing the frequency vector~$\hat{\fp}{}^{(d)}$ from the given OPs $\uppi_1,\ldots,\uppi_n$ with $n\in \bbn$. Then, the scaled deviations $\sqrt{n}\, \big(\hat{\fp}{}^{(d)}-\fp_0\big)$ from uniformity, where $\fp_0=(1/\cat!,\ldots,1/\cat!)$, are asymptotically normally distributed according to $\norm(\0, \fSigma_\cat)$ under the \iid-null, see \citet{weiss22} for details, and this can be used for developing nonparametric dependence tests. 

\smallskip
Before providing more details on these tests, two difficulties have to be mentioned. First, the number of different OPs increases with~$\cat!$ and thus quickly becomes unfeasibly large. Already for $\cat=4$, there are 24~different OPs such that reasonable frequencies are hardly obtained for typical time-series lengths~$n$. On the other hand, the choice $\cat=2$ leads to a rather crude discretization and a poor performance of the resulting OP-tests \citep{weiss22}. Thus, in the sequel, we follow the recommendation of \citet{bandt19} and solely focus on OPs of order $\cat=3$. These are given (in lexicographic ordering) by
\begin{equation}
\label{OrdPatt3}
\begin{array}{@{}cccccc@{}}
(1,2,3) & (1,3,2) & (2,1,3) & (2,3,1) & (3,1,2) & (3,2,1) \\
\includegraphics[viewport=75 85 135 145, clip=, scale=0.65]{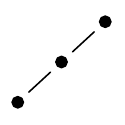}
 & \includegraphics[viewport=75 85 135 145, clip=, scale=0.65]{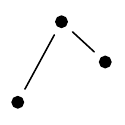} 
 & \includegraphics[viewport=75 85 135 145, clip=, scale=0.65]{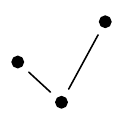} 
 & \includegraphics[viewport=75 85 135 145, clip=, scale=0.65]{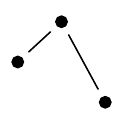} 
 & \includegraphics[viewport=75 85 135 145, clip=, scale=0.65]{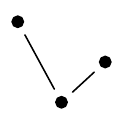} 
 & \includegraphics[viewport=75 85 135 145, clip=, scale=0.65]{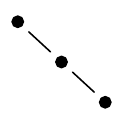} 
\end{array}
\end{equation}
The second difficulty arises from the fact that even if $(X_t)$ is \iid, the resulting OP~series $(\uppi_t)$ is not serially independent but exhibits a kind of $(\cat-1)d$-dependence. This complicates the computation of the covariance matrix~$\fSigma_\cat$ of $\sqrt{n}\, \big(\hat{\fp}{}^{(d)}-\fp_0\big)$ according to $\norm(\0, \fSigma_\cat)$. But again, for $\cat=3$, a closed-form expression for~$\fSigma_\cat$ is known \citep{weiss22}. We therefore proceed by summarizing the corresponding dependence tests.

\smallskip
The first group of dependence tests refers to the three entropy-like statistics
\ba
\label{entropies}
\begin{array}{l}
\text{entropy:}\qquad 
\widehat{H}^{(d)} = H\big(\hat{\fp}^{(d)}\big)\ =\ 
-\sum\limits_{i=1}^{\cat!} \hat{p}_i^{(d)}\,\ln{\hat{p}_i^{(d)}},
\\[1ex]
\text{extropy:}\qquad
\widehat{H}_{\textup{ex}}^{(d)} = H_{\textup{ex}}\big(\hat{\fp}^{(d)}\big)
\ =\ 
-\sum\limits_{i=1}^{\cat!} \big(1-p_i^{(d)}\big)\,\ln\big(1-p_i^{(d)}\big),
\\[1ex]
\text{distance to white noise:}\qquad
\widehat{\Delta}_2^{(d)} = \Delta_2\big(\hat{\fp}^{(d)}\big)\ =\ \sum\limits_{i=1}^{\cat!} \big(\hat{p}_i^{(d)}-1/\cat!\big)^2,
\end{array}
\ea
which all take nonnegative values (with increasing values for increasing deviation from uniformity). Note that the statistics \eqref{entropies} are indeed well defined for any order $\cat\geq 2$, not just for $\cat=3$. As shown by \citet{weiss22}, the following results holds.

\begin{prop}
\label{propEntropies}
Let $(X_t)_\bbz$ be an \iid\ real-valued and continuously distributed process, let $\cat=3$. Then,
$$
\textstyle
n\,\widehat{\Delta}_2^{(d)},
\quad 
-\frac{n}{3}\,\Big(\widehat{H}^{(d)}-\ln{6}\Big),
\quad 
-\frac{5n}{3}\,\Big(\widehat{H}_{\textup{ex}}^{(d)}-5\,\ln\big(\frac{6}{5}\big)\Big)
$$
are asymptotically distributed like the quadratic form
$$
\textstyle
Q_3\ =\ \frac{1}{12} (2+\sqrt{2})\cdot \chi_1^2\ +\ \frac{2}{15}\cdot \chi_1^2\ +\ \frac{1}{10}\cdot \chi_1^2\ +\ \frac{1}{12}(2-\sqrt{2})\cdot \chi_1^2.
$$
\end{prop}
The resulting three entropy-like tests of the \iid-null can be implemented in practice by computing either P-values or critical values with respect to the quadratic-form (QF) distribution of~$Q_3$, which is possible by using the R~package \href{https://cran.r-project.org/package=CompQuadForm}{\nolinkurl{CompQuadForm}} of \citet{duchesne10}.

\smallskip
In the case $\cat=3$, a second group of OP-statistics is defined by linear expressions of the form $\fa\!^\top\,(\hat{\fp}{}^{(d)}-\fp_0)$ with fixed coefficient vector $\fa\in\bbr^6$. More specifically, \citet{bandt19} proposed the four statistics
\ba
\label{LinStat3}
\begin{array}{ll}
\text{up-down balance:} & 
\hat{\upbeta}^{(d)} = \hat{p}_1^{(d)} - \hat{p}_6^{(d)},\\[1ex]
\text{persistence:} & 
\hat{\uptau}^{(d)} = \hat{p}_1^{(d)} + \hat{p}_6^{(d)}-\frac{1}{3},\\[1ex]
\text{rotational asymmetry:} & 
\hat{\upgamma}^{(d)} = \hat{p}_3^{(d)} + \hat{p}_4^{(d)} - \hat{p}_2^{(d)} - \hat{p}_5^{(d)},\\[1ex]
\text{up-down scaling:} & 
\hat{\updelta}^{(d)} = \hat{p}_2^{(d)} + \hat{p}_3^{(d)} - \hat{p}_4^{(d)} - \hat{p}_5^{(d)}.
\end{array}
\ea
\citet{weiss22} derived the following normal asymptotics.

\begin{prop}
\label{propLinearStatistics}
Let $(X_t)_\bbz$ be an \iid\ real-valued and continuously distributed process, let $\cat=3$. Then,
$$
\begin{array}{l@{\qquad}l}
\sqrt{n}\,\hat{\upbeta}^{(d)}\ \overset{\text{a}}{\sim}\ \norm\big(0,\ 1/3\big),
&
\sqrt{n}\,\hat{\uptau}^{(d)}\ \overset{\text{a}}{\sim}\ \norm\big(0,\ 8/45\big),
\\[2ex]
\sqrt{n}\,\hat{\upgamma}^{(d)}\ \overset{\text{a}}{\sim}\ \norm\big(0,\ 2/5\big),
&
\sqrt{n}\,\hat{\updelta}^{(d)}\ \overset{\text{a}}{\sim}\ \norm\big(0,\ 2/3\big).
\end{array}
$$
\end{prop}
In Section~\ref{The Novel Hilbert-Curve Approach} below, we propose to combine the dependence tests of Propositions~\ref{propEntropies} and~\ref{propLinearStatistics} with a Hilbert-curve approach to test for spatial dependence on regular grid data. In the particular case of 2D data, \ie if the data are given as an $(n_1+1)\times (n_2+1)$-rectangle $(x_{\fti}) = (x_{t_1,t_2})$ with $0\leq t_1\leq n_1$ and $0\leq t_2\leq n_2$, a tailor-made family of spatial-dependence tests is already available, which shall serve as competitors in the performance analyses of Section~\ref{Performance Analyses}. These tests are based on SOPs, which are computed from $(x_{\fti})$ by successively mapping the $2\times 2$-squares
\begin{equation}
\label{SOPs}
%\mX_{\fti} = 
\left(\begin{array}{cc}
x_{t_1-1,t_2-1} & x_{t_1-1,t_2} \\
x_{t_1,t_2-1} & x_{t_1,t_2} \\
\end{array}\right) =: \left(\begin{array}{cc}
x_1 & x_2 \\
x_3 & x_4 \\
\end{array}\right)
\quad\text{onto}\quad
\left(\begin{array}{cc}
r_1 & r_2 \\
r_3 & r_4\\
\end{array}\right),
\end{equation}
where $(r_1, r_2, r_3, r_4)\in S_4$ is the OP \eqref{ordpatternrank} corresponding to $(x_1, x_2, x_3, x_4)$. For further details on SOPs and the proposed spatial-dependence tests, we refer the reader to \citet{bandtw23,WeissKim2024,WeissKim2025}.

\subsection{The Novel Hilbert-Curve Approach}
\label{The Novel Hilbert-Curve Approach}
From now on, the considered DGP is a $k$-dimen\-sional, real-valued and continuously distributed random field $(X_{\fti})=(X_{\fti})_{\fti\in\bbz^k}$. The realized data sets are kinds of cuboids, $(x_{\fti})=(x_{\fti})_{\fti\in\mathcal{T}}$ with data dimension $\mathcal{T}=\{0,\ldots,n_1\}\times\cdots\times \{0,\ldots,n_k\}$ of cardinality $|\mathcal{T}| = (n_1+1)\cdots (n_k+1)$. For example, we are concerned with data rectangles if $k=2$ (``2D case'', recall the discussion around \eqref{SOPs}), and with true cuboids if $k=3$ (``3D case''). Our task is to test the null hypothesis that $(X_{\fti})$ is \iid {} For this purpose, we propose the following \textbf{Hilbert-curve approach:}

\begin{enumerate}
	\item Compute a $k$-dimensional Hilbert curve corresponding to the data dimension $\mathcal{T}$ of cardinality $|\mathcal{T}|$. It can be expressed as a sequence $\mathcal{H}=(\ftau_t)_{t=1,\ldots, |\mathcal{T}|}$ of $k$-dimensional vectors, which constitutes an enumeration of the elements in~$\mathcal{T}$.
	
	\item Transform the grid data $(x_{\fti})_{\mathcal{T}}$ into the sequence $(x_{\ftaui_t})_{\mathcal{H}}$ corresponding to the Hilbert curve~$\mathcal{H}$, and apply a serial-dependence test from Proposition~\ref{propEntropies} or~\ref{propLinearStatistics}.
\end{enumerate}
Recall that \citet{chagas21,bariviera25} used an analogous Hilbert-curve ($\mathcal{H}$) transformation for texture analysis of 2D image data. 
Note that under the \iid-null, the proposed $\mathcal{H}$-rearrangement of the data does not affect the \iid-property, \ie $(X_{\ftaui_t})_{\mathcal{H}}$ is an \iid\ sequence such that the asymptotics from Propositions~\ref{propEntropies} and~\ref{propLinearStatistics} remain valid. In particular, the finite sample performance of the sizes (\ie rates of false rejections of the null), as analyzed in \citet{weiss22}, does not change. So we conclude that even for rather small data sets such as $|\mathcal{T}|=100$, we have a sufficiently close agreement between the nominal level of the tests and their actual size. Therefore, our performance analyses in Section~\ref{Performance Analyses} focus solely on the power properties of the Hilbert-curve OP-tests (``$\mathcal{H}$OP-tests'' for short).

\begin{figure}[t]
\centering\footnotesize
\begin{tabular}{cccc}
$(10,10)$ & $(15,15)$ & $(20,20)$ & $(40,25)$ \\[1ex]
\includegraphics[viewport=5 5 125 125, clip=, scale=0.65]{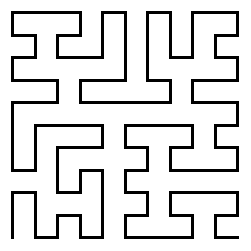}
&
\includegraphics[viewport=5 5 125 125, clip=, scale=0.65]{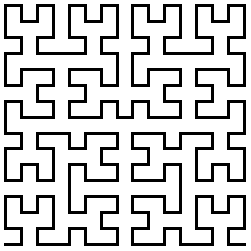}
&
\includegraphics[viewport=5 5 125 125, clip=, scale=0.65]{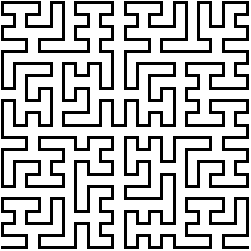}
&
\includegraphics[viewport=5 5 125 125, clip=, scale=0.65]{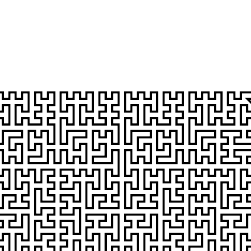}
\\
\\[2ex]
\includegraphics[viewport=5 5 125 125, clip=, scale=0.65]{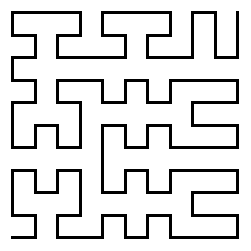}
&
\includegraphics[viewport=5 5 125 125, clip=, scale=0.65]{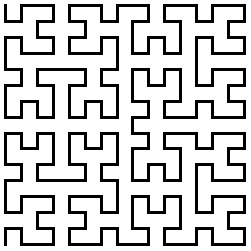}
&
\includegraphics[viewport=5 5 125 125, clip=, scale=0.65]{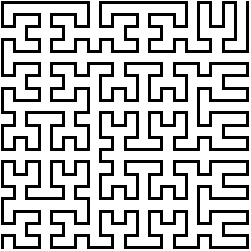}
&
\includegraphics[viewport=5 5 125 125, clip=, scale=0.65]{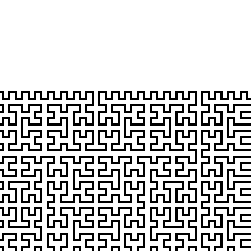}
\end{tabular}
\caption{2D Hilbert curves for simulation study in Section~\ref{Performance Analyses} with data dimensions $(n_1,n_2)$. First row computed by \href{https://github.com/jakubcerveny/gilbert}{\nolinkurl{gilbert}}, second by \citet{rong21}.}
\label{figHilbert2D}
\end{figure}

\smallskip
Due to their relevance in real-world applications, our subsequent performance analyses concentrate on the 2D and 3D case. As already mentioned in Section~\ref{Introduction}, 2D Hilbert curves for arbitrary-sized rectangles can be computed by using the algorithm of \citet{rong21} with corresponding Matlab implementation, or the algorithm from the Github project \href{https://github.com/jakubcerveny/gilbert}{\nolinkurl{gilbert}} with codes in Python, C, etc., and with additional \texttt{Julia} implementation \href{https://github.com/CliMA/GilbertCurves.jl}{\nolinkurl{GilbertCurves}} in the 2D case (all software implementations accessed on May 26, 2025). It should be noted, however, that these algorithms usually lead to different Hilbert curves. The algorithms of \citet{rong21} and \href{https://github.com/jakubcerveny/gilbert}{\nolinkurl{gilbert}} can also be used to compute 3D Hilbert curves as required for data cuboids. However, one has to note that the algorithm of \citet{rong21} requires the third data dimension to have a length of a power of two.
The command \href{https://reference.wolfram.com/language/ref/HilbertCurve.html}{\nolinkurl{HilbertCurve}} in Wolfram Mathematica is also worth mentioning, since it allows for arbitrary dimension $k\geq 2$, but its application is limited to cubes with the length being a power of two.

\begin{figure}[H]
\centering\footnotesize
\begin{tabular}{cccc}
$(7,7,7)$ & $(10,10,7)$ & $(10,15,7)$ & $(15,15,7)$ \\[1ex]
\includegraphics[viewport=30 30 240 240, clip=, scale=0.35]{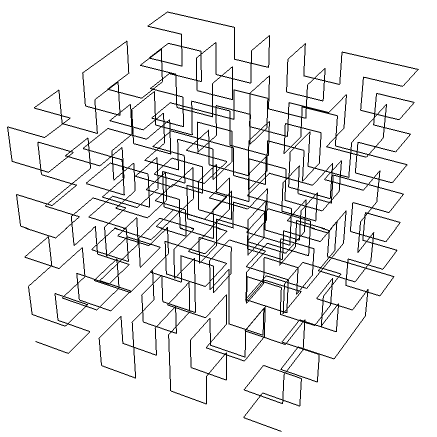}
&
\includegraphics[viewport=30 30 240 240, clip=, scale=0.35]{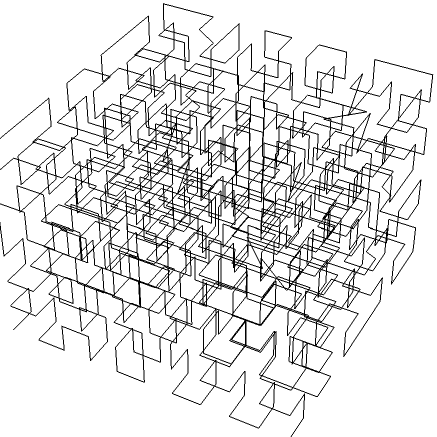}
&
\includegraphics[viewport=30 30 240 240, clip=, scale=0.35]{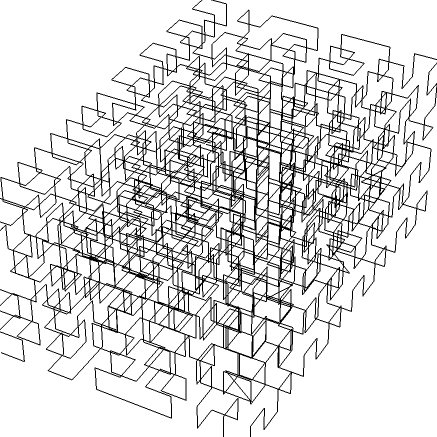}
&
\includegraphics[viewport=30 30 240 240, clip=, scale=0.35]{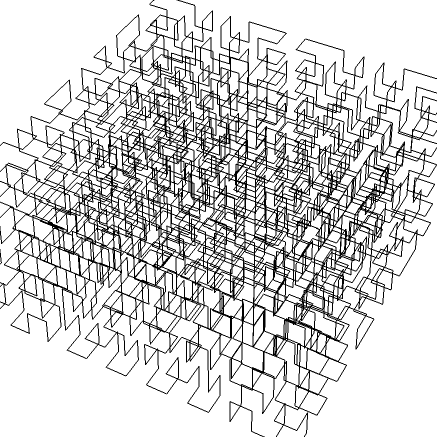}
\\[2ex]
\includegraphics[viewport=30 30 240 240, clip=, scale=0.35]{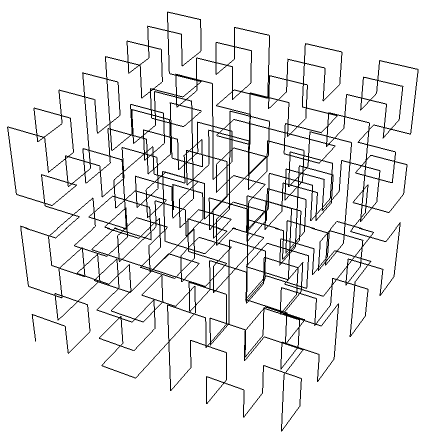}
&
\includegraphics[viewport=30 30 240 240, clip=, scale=0.35]{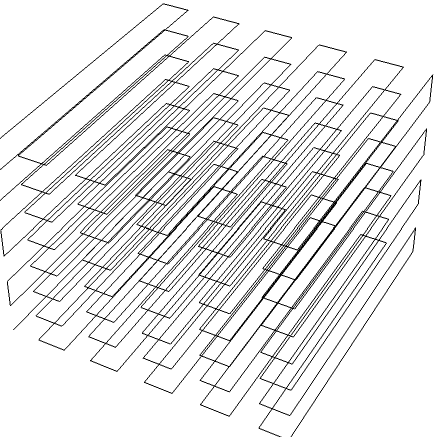}
&
\includegraphics[viewport=30 30 240 240, clip=, scale=0.35]{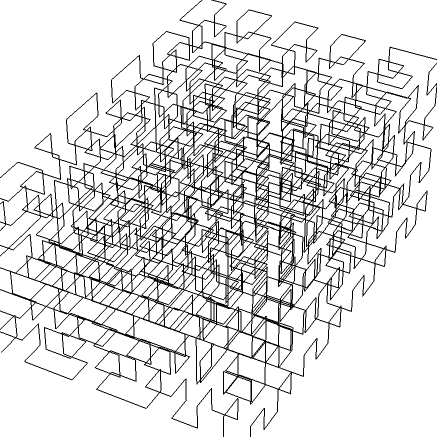}
&
\includegraphics[viewport=30 30 240 240, clip=, scale=0.35]{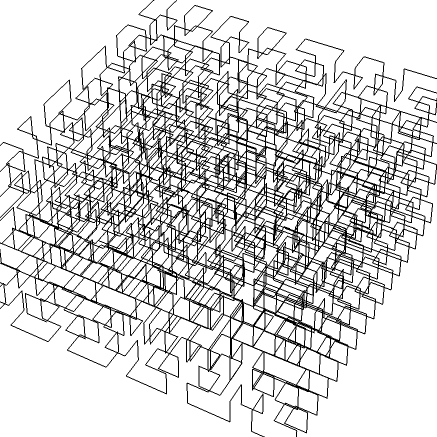}
\\
\end{tabular}
\caption{3D Hilbert curves for simulation study in Section~\ref{Performance Analyses} with data dimensions $(n_1,n_2,n_3)$. First row computed by \href{https://github.com/jakubcerveny/gilbert}{\nolinkurl{gilbert}}, second by \citet{rong21}.}
\label{figHilbert3D}
\end{figure}

\smallskip
We emphasize that the definition of a (generalized) Hilbert curve is not unique. In fact, the aforementioned algorithms by \href{https://github.com/jakubcerveny/gilbert}{\nolinkurl{gilbert}} and \citet{rong21} usually lead to different Hilbert curves, which, in turn, affects the value of the computed $\mathcal{H}$OP-statistic. This is illustrated by Figures~\ref{figHilbert2D} and~\ref{figHilbert3D}, where Hilbert curves by both algorithms are shown for the data dimensions considered in our simulation study in Section~\ref{Performance Analyses} below. A particularly strong difference can be noted in the 3D scenario $(n_1,n_2,n_3) = (10,10,7)$, with the curve by \citet{rong21} looking quite unusual, whereas the remaining Hilbert curves have a similar appearance. Since the actual value of the $\mathcal{H}$OP-statistic depends on the chosen Hilbert curve, it is reasonable to ask how this choice affects the power performance of the $\mathcal{H}$OP-tests. This is one of the main research questions for our performance analyses in Section~\ref{Performance Analyses}.

\numberwithin{satz}{section}

\section{Performance Analyses}
\label{Performance Analyses}
We assess the effectiveness of our proposed $\mathcal{H}$OP-tests, along with those of relevant benchmark methods, by evaluating their true rejection rates against the \iid\ assumption across a range of simulated DGPs. 
Under the null hypothesis of spatial independence, the proposed Hilbert curves transform the data into an \iid\ sequence, regardless of the dimensionality of the original data and the chosen Hilbert curve. Consequently, the size properties of our $\mathcal{H}$OP-tests are comparable to those analyzed by \citet{weiss22}, who demonstrated close agreement with the nominal level even with modest sample sizes. To avoid redundancies, we omit an investigation of the $\mathcal{H}$OP-tests' sizes, but will focus solely on their power. We used the \texttt{Julia} package \href{https://github.com/AdaemmerP/OrdinalPatterns.jl}{\nolinkurl{OrdinalPatterns.jl}} for the simulations.

\subsection{Power Analysis for 2D Scenarios}
\label{Power Analysis for 2D Scenarios}
We consider the continuous DGPs from \citet{WeissKim2024} and also apply part of their used competitors. We emphasize that we do not expect our new $\mathcal{H}$OP tests to outperform the SOP tests in the 2D case, since SOPs, as those investigated by \citet{WeissKim2024}, were specifically designed for 2D grids. However, the advantage of our $\mathcal{H}$OP approaches is that they can be generally applied to 2D, 3D and even higher dimensions. Therefore, in this subsection, our focus is on exploring whether the $\mathcal{H}$OP tests can compete with the SOP approaches in the 2D case (while they are unrivaled for 3D and higher dimensions), rather than on whether they outperform them. \citet{WeissKim2024} analyzed the performance of various SOP-based tests and found that the so-called ``${\widetilde{\tau}}$-test'', which should not be confused with our $\mathcal{H}$OP-based ``${\hat{\uptau}}$-test'' \eqref{LinStat3}, uniquely outperformed all other SOP-tests. Therefore, we use $\widetilde{\tau}$ as our first competitor in the 2D case. It computes the relative frequency of observing SOPs from the set
$$
\mathcal S_3\ =\  \Big\{\left(\begin{smallmatrix} 1 & 3 \\ 4 & 2 \end{smallmatrix}\right),
\left(\begin{smallmatrix} 1 & 4 \\ 3 & 2 \end{smallmatrix}\right),
\left(\begin{smallmatrix} 2 & 3 \\ 4 & 1 \end{smallmatrix}\right),
\left(\begin{smallmatrix} 2 & 4 \\ 3 & 1 \end{smallmatrix}\right),
\left(\begin{smallmatrix} 3 & 1 \\ 2 & 4 \end{smallmatrix}\right),
\left(\begin{smallmatrix} 3 & 2 \\ 1 & 4 \end{smallmatrix}\right),
\left(\begin{smallmatrix} 4 & 1 \\ 2 & 3 \end{smallmatrix}\right),
\left(\begin{smallmatrix} 4 & 2 \\ 1 & 3 \end{smallmatrix}\right)\Big\},
$$
the so-called ``type-3 SOPs''. The ${\widetilde{\tau}}$-statistic subtracts the value~$1/3$ from the relative frequency of $\mathcal{S}_3$, corresponding to the null probability for observing a SOP from~$\mathcal S_3$. Our second competitor is the spatial autocorrelation function (SACF), defined by 
$$
\hat\rho(\fh)\ =\ \frac{\sum_{\fti\in\mathcal{T}} (X_{\fti}-\overline{X})(X_{\fti-\fhi}-\overline{X})}{\sum_{\fti\in\mathcal{T}} (X_{\fti}-\overline{X})^2}
\quad\text{with spatial lag }
\fh\in\bbz^k,
$$
where~$\overline{X} = \frac{1}{|\mathcal{T}|}\,\sum_{\fti\in\mathcal{T}} X_{\fti}$. The critical values for the $\hat{\rho}(\1)$-test are quantiles from a normal distribution with mean~$0$ and variance $1/|\mathcal{T}|$, see \citet{meyer17} for details. \citet{WeissKim2024} used the $\hat{\rho}(\1)$-test as a competitor in their study of 2D random fields.

\smallskip
For all simulations involving 2D and 3D grids, we found  that using $\hat{\uptau}$ in \eqref{LinStat3} produced the highest overall power, whereas the power for $\hat{\upbeta}$, $\hat{\upgamma}$ and $\hat{\updelta}$ in \eqref{LinStat3} was very low in any case. The entropy-like statistics in \eqref{entropies}, by contrast, turned out to be powerful and produced quite similar results among each other. Hence, for conciseness and better comparability, we only present, for the $\mathcal{H}$OP approaches, the power results of the ``$\hat{\uptau}$-test'' and entropy ``$\widehat{H}$-test''. Our findings for the 2D case are in close agreement to those of \citet{weiss22} concerning time-series data, who also found the highest power for $\hat{\uptau}$ and a similar power for all entropy-like statistics. 
For comparison, we use the same grid sizes as in \citet{WeissKim2024}, namely $(n_1,n_2)\in\big\{(10,10), (15,15), (20,20), (40,25)\big\}$ corresponding to the cardinalities $|\mathcal{T}|\in \{121, 256, 441, 1066\}$, and we report the power values of the ${\widetilde{\tau}}$-test and the $\hat{\rho}(\1)$-test as computed by \citet{WeissKim2024} (always in italic font).
For both $\hat{\uptau}$ and $\widehat{H}$, however, we consider the delays $d = 1, \ldots, 4$, because Hilbert curves move nonlinearly through the space, which may contribute to variations in power across different delays.

\smallskip
We start our power analyses with the linear first-order spatial autoregressive (SAR$(1,1)$) process 
\begin{equation} \label{SAR11}
       X_{t_{1}, t_{2}}= \alpha_1\cdot X_{{t_1}-1, t_{2}} + \alpha_2\cdot  X_{t_{1}, t_{2}-1}+ \alpha_3\cdot  X_{t_{1}-1, t_{2}-1} + \varepsilon_{t_{1}, t_{2}},
\end{equation}
    where $\varepsilon_{t_{1}, t_{2}}$ are \iid\ $\norm(0,1)$.
We consider the parameterizations $(\alpha_1, \alpha_2, \alpha_3)$ equal to $(0.2, 0.2, 0.2)$ and $(0.4, 0.3, 0.1)$, which we refer to as DGPs ``2D-1.1'' and ``2D-1.2'', respectively. The results are shown in Table~\ref{tab_sar11}. Regardless of the grid size, the SACF has the highest rejection rate among all approaches. Given the linear dependence of the SAR$(1,1)$ process, the strong performance of the linear and parametric $\hat{\rho}(\1)$-test is not surprising. Among the $\mathcal{H}$OP-statistics, the power of the $\hat{\uptau}$-tests is highest. 
When comparing the rejection rates of the (SOP) $\widetilde{\tau}$-test with those of the ($\mathcal{H}$OP) $\hat{\uptau}$-test, the $\mathcal{H}$OP-statistic $\hat{\uptau}$ is highly competitive with respect to power. Another interesting finding is that, in some cases, the power of the $\mathcal{H}$OP begins to increase again at greater delays. For example, the power of $\hat{\uptau}$ and $\widehat{H}$ doubles from delay 2 to 3 for grid sizes $(40, 25)$ and DGP ``2D-1.2''. Furthermore, there appears no substantial difference in the power values whether the $\mathcal{H}$OP tests depend on the Hilbert curves by \href{https://github.com/jakubcerveny/gilbert}{\nolinkurl{gilbert}} or \citet{rong21}.

\smallskip
Contaminating the observations from the SAR$(1,1)$ process in \eqref{SAR11} with additive outliers (AOs)---by which 10\% of the data are randomly increased by~$+10$---produces the results presented in Table~\ref{tab_sar11_ao}. These contaminated DGPs are labeled as ``2D-2.1'' and ``2D-2.2'', respectively. In contrast to Table~\ref{tab_sar11}, the power of the nonparametric SOP- and $\mathcal{H}$OP-tests is now much higher than that of $\hat{\rho}(\1)$. Again, $\hat{\uptau}$ performs overall better than the classical entropy measure, $\widehat{H}$, and $\hat{\uptau}$ is also competitive to the SOP-approach $\widetilde{\tau}$.  

\smallskip
Next, we consider the multilateral first-order simultaneous AR (SAR$(1)$) model defined by
\ba
\label{simultAR1}
X_{t_{1},t_{2}} = a_{1} \cdot X_{t_{1}-1,t_{2}} +   a_{2} \cdot X_{t_{1},t_{2}-1} +  a_{3} \cdot X_{t_{1},t_{2}+1} +  a_{4}\cdot X_{t_{1}+1,t_{2}} +  \varepsilon_{t_{1}, t_{2}},
\ea
where $\varepsilon_{t_{1}, t_{2}}$ are \iid\ $\norm(0,1)$. We consider the isotropic DGP with $a_1=\cdots=a_4=0.1$ (abbreviated by ``2D-3.1''), and the anisotropic DGP with $a_1 = a_2 = 0.05$ and $a_3=a_4= 0.15$ (abbreviated by ``2D-3.2''). The results are shown in Table~\ref{tab_sar1}. In these scenarios, the $\hat{\rho}(\1)$-test performs poorly compared to the SOP- and $\mathcal{H}$OP-approaches. Although the SOP $\widetilde{\tau}$-test outperforms the $\mathcal{H}$OP-based tests, in particular for larger grid sizes, $\hat{\uptau}$ remains again competitive. Contaminating the values of the SAR(1) process with AOs (10\% of values have added value $+5$) yields the results in Table~\ref{tab_sar1_ao}. The DGPs are denoted as ``2D-4.1'' and ``2D-4.2'', depending on the parameter choices as in Table~\ref{tab_sar1}. Note that for the second parameterization, we corrected a typo in the power values for the $\hat{\rho}(\1)$-test in \citet{WeissKim2024}. Although the power for all test statistics decreases overall, the nonparametric SOP- and $\mathcal{H}$OP-approaches are clearly less sensitive to additive outliers than the parametric $\hat{\rho}(\1)$-test, resulting in much higher rejection rates for the SOP- and $\mathcal{H}$OP-tests than for the $\hat{\rho}(\1)$-test when dealing with AOs. Again, the results do not appear to depend on whether the Hilbert curves are calculated by either \href{https://github.com/jakubcerveny/gilbert}{\nolinkurl{gilbert}} or \citet{rong21}. 

\smallskip
Next, we consider the unilateral but nonlinear first-order spatial quadratic MA (SQMA$(1,1)$) process with \iid\ $\norm(0,1)$-innovations~$\varepsilon_{t_{1}, t_{2}}$, where
\ba
\label{SQMA11}
X_{t_{1}, t_{2}}= \beta_{1} \cdot \varepsilon_{{t_1}-1, t_{2}}^{q_1} + \beta_{2} \cdot  \varepsilon_{t_{1}, t_{2}-1}^{q_2}+ \beta_{3} \cdot  \varepsilon_{t_{1}-1, t_{2}-1}^{q_3} + \varepsilon_{t_{1}, t_{2}},
\ea
with $\beta_1=\cdots=\beta_3=0.8$. We consider exponents $(q_1,q_2,q_3) = (2,2,2)$ for DGP ``2D-5.1'' and exponents $(2,1,2)$ for DGP ``2D-5.2'', see Table~\ref{tab_sqma11}. Overall, the results highlight again the superiority of the nonparametric SOP- and $\mathcal{H}$OP-approaches over the parametric $\hat{\rho}(\1)$-test. Among the $\mathcal{H}$OP-tests, $\hat{\uptau}$ outperforms the classical entropy approach, $\widehat{H}$. Again, it appears irrelevant whether the Hilbert curves are calculated by \href{https://github.com/jakubcerveny/gilbert}{\nolinkurl{gilbert}} or \citet{rong21}. If the power of a test based on \href{https://github.com/jakubcerveny/gilbert}{\nolinkurl{gilbert}} is higher than that based on \citet{rong21}, the results are reverse for higher delays (\eg delays one and two for DGP ``2D-5.1'' with $(40, 25)$). Furthermore, the power of the $\mathcal{H}$OP-tests begins to increase again with greater delay.

Finally, we consider the multilateral extension of SQMA$(1, 1)$, namely the first-order simultaneous QMA (SQMA$(1)$) process defined by
\ba
\label{BSQMA11}
X_{t_{1}, t_{2}}= b_{1} \cdot \varepsilon_{{t_1}-1, t_{2}-1}^{q_1} + b_{2} \cdot \varepsilon_{t_{1}+1, t_{2}-1}^{q_2} + b_{3}\cdot \varepsilon_{t_{1}+1, t_{2}+1}^{q_3} + b_{4} \cdot \varepsilon_{t_{1}-1, t_{2}+1}^{q_4} + \varepsilon_{t_{1}, t_{2}}
\ea
with $b_1=\cdots=b_4=0.8$. For these scenarios, we choose the exponents $(2,\ldots,2)$ for DGP ``2D-6.1'', and $(2,1,2,1)$ for DGP ``2D-6.2''. The results are presented in Table~\ref{tab_bsqma11}. In general, compared to Table~\ref{tab_sqma11}, the rejections rates of the SOP- and $\mathcal{H}$OP-approaches decrease, whereas the power of the SACF increases. Nevertheless, the power of $\hat{\rho}(\1)$ is still lower than those of the nonparametric approaches in all cases. Comparing the nonparametric approaches with each other shows that, in several cases (see e.g. DGP ``2D-6.1'' with grid sizes $(20, 20)$ and $(40, 25)$), the $\mathcal{H}$OP-test $\hat{\uptau}$ now even outperforms the SOP-test $\widetilde{\tau}$. As before, the results do not appear to depend on whether the Hilbert curves are calculated by either \href{https://github.com/jakubcerveny/gilbert}{\nolinkurl{gilbert}} or \citet{rong21}.  Apart from that, $\hat{\uptau}$ is more effective than $\widehat{H}$ in detecting spatial dependence and competitive with the (SOP) $\widetilde{\tau}$-test. Furthermore, the rejection rates for our $\mathcal{H}$OP-tests begin to increase again at higher delays in several cases.

\bigskip
Let us sum up the 2D power analyses. As explained in the beginning of Section~\ref{Power Analysis for 2D Scenarios}, there already exists a tailor-made and OP-based solution for testing for spatial dependence in 2D grid data, namely SOP-based approaches like the $\widetilde{\tau}$-test. Thus, we did not anticipate to outperform the SOP-tests with our universally applicable $\mathcal{H}$OP approaches, but the SOP-tests allow to calibrate the performance of the $\mathcal{H}$OP approaches. 
Here, the power analyses for the 2D scenarios indeed show that, although the SOP $\widetilde{\tau}$-test is generally superior, our proposed $\hat{\tau}$-test is highly competitive and strongest among all $\mathcal{H}$OP-tests. In addition, we did not see substantial differences of whether the Hilbert curves are calculated based on the approach by \href{https://github.com/jakubcerveny/gilbert}{\nolinkurl{gilbert}} or \citet{rong21}. This is an important finding for practice, because Hilbert curves are not uniquely defined and may differ for different software implementations. The parametric $\hat{\rho}(\1)$-test achieves the highest rejections rates solely under linear and unilateral DGPs, but its effectiveness diminishes substantially in the presence of outliers and nonlinear DGPs. These findings coincide with those in \citet{weiss22} and \citet{WeissKim2024}. Interestingly, and in contrast to previous findings, the $\mathcal{H}$OP-tests demonstrate improved efficiency at larger delays. This underscores the possible advantage of an approach being capable of aggregating information across various delays, suggesting a promising avenue for future research.

\subsection{Power Analysis for 3D Scenarios}
\label{Power Analysis for 3D Scenarios}
Unlike for 2D random fields, no (S)OP-tests have yet been developed for the 3D case so far, which underscores the novelty and relevance of our suggested $\mathcal{H}$OP-approach. For this reason, we also only use the 3D $\hat{\rho}(\1)$-test as a competitor in the following power analyses. Furthermore, we newly had to define relevant 3D power scenarios, where we tried to reach a high comparability to the 2D scenarios from Section~\ref{Power Analysis for 2D Scenarios}. 
Therefore, we defined the grids $(n_1,n_2,n_3)\in\big\{(7,7,7), (10,10,7), (10,15,7), (15,15,7)\big\}$, which lead to the cardinalities $|\mathcal{T}|\in \{512, 968, 1408, 2048\}$ being quite similar to those in Section~\ref{Power Analysis for 2D Scenarios}. We also defined analogous 3D DGPs, although modifications are needed in some cases, because the 3D power scenarios are computationally much more demanding than the 2D ones.

\smallskip
As MA-like models can efficiently be implemented in any dimension, we first consider direct 3D counterparts to \eqref{SQMA11} and \eqref{BSQMA11}. The unilateral SQMA$(1,1,1)$ DGP is defined as
\begin{align}
\label{SQMA111}
X_{t_{1}, t_{2}, t_{3}}= &\ \beta_{1} \cdot \varepsilon_{{t_1}-1, t_{2}, t_{3}}^{q_1} + \beta_{2} \cdot  \varepsilon_{t_{1}, t_{2}-1, t_{3}}^{q_2}+ \beta_{3} \cdot  \varepsilon_{t_{1}, t_{2}, t_{3}-1}^{q_3}\\
\nonumber
& + \beta_{4} \cdot  \varepsilon_{t_{1}-1, t_{2}-1, t_{3}}^{q_4} + \beta_{5} \cdot  \varepsilon_{t_{1}-1, t_{2}, t_{3}-1}^{q_5} + \beta_{6} \cdot  \varepsilon_{t_{1}, t_{2}-1, t_{3}-1}^{q_6}\\
\nonumber
& + \beta_{7} \cdot  \varepsilon_{t_{1}-1, t_{2}-1, t_{3}-1}^{q_7} + \varepsilon_{t_{1}, t_{2}, t_{3}},
\end{align}
with $\beta_1=\cdots=\beta_7=0.8$. We consider exponents $(q_1,\ldots,q_7) = (2,\ldots,2)$ for DGP ``3D-1.1'' and $(2, 2, 2, 1, 1, 1, 2)$ for DGP ``3D-1.2''. Results are shown in Table~\ref{tab_sqma111}. Here, the nonparametric $\mathcal{H}$OP-approaches generally have much higher rejection rates than the $\hat{\rho}(\1)$-test, particularly at delay 1. Also, the power of the $\hat{\uptau}$-test is always higher than that of the entropy approach $\widehat{H}$. An important point is the effect of the actual choice for the Hilbert curve. Recall that for our 2D scenarios, we did not observe any notable differences in power performance. In the 3D case, however, this is not always true anymore. For the grid size $(10, 10, 7)$, the Hilbert curves computed using the approaches by \href{https://github.com/jakubcerveny/gilbert}{\nolinkurl{gilbert}} and \citet{rong21} differ substantially, as can be seen in Figure~\ref{figHilbert3D}. In fact, the Hilbert curve by \citet{rong21} looks quite unusual such that one can doubt if the aforementioned ``clustering property'' (\ie the property of preserving locality) still holds. As a result, the power profile of the corresponding $\mathcal{H}$OP-tests differs considerably in this particular case. For the remaining grid sizes, however, it does not make much difference how the Hilbert curves are computed. Another finding that mirrors our results for the 2D case is that, in several scenarios, the power increases again with higher delays, though it never exceeds the power at delay 1 (\eg ``3D-1.1'' with $(10, 15, 7)$, and ``3D-1.2'' with $(15, 15, 7)$). 

\smallskip
Next, we consider the multilateral SQMA$(1)$ process, defined by
\begin{align}
\label{BSQMA111}
X_{t_{1}, t_{2}, t_{3}}=&\ b_{1} \cdot \varepsilon_{{t_1}-1, t_{2}-1, t_{3}-1}^{q_1} + b_{2} \cdot \varepsilon_{t_{1}+1, t_{2}-1, t_{3}-1}^{q_2} + b_{3}\cdot \varepsilon_{t_{1}-1, t_{2}+1, t_{3}-1}^{q_3}\\
\nonumber
& + b_{4} \cdot \varepsilon_{t_{1}-1, t_{2}-1, t_{3}+1}^{q_4} + b_{5} \cdot \varepsilon_{t_{1}+1, t_{2}+1, t_{3}-1}^{q_5} + b_{6} \cdot \varepsilon_{t_{1}+1, t_{2}-1, t_{3}+1}^{q_6}\\
\nonumber
& + b_{7} \cdot \varepsilon_{t_{1}-1, t_{2}+1, t_{3}+1}^{q_7} + b_{8} \cdot \varepsilon_{t_{1}+1, t_{2}+1, t_{3}+1}^{q_8} + \varepsilon_{t_{1}, t_{2}, t_{3}}
\end{align}
with $b_1=\cdots=b_8=0.8$. We consider exponents $(q_1,\ldots,q_8) = (2,\ldots,2)$ for DGP ``3D-2.1'' and $(2, 2, 1, 2, 1, 2, 1, 2)$ for DGP ``3D-2.2''. The results are shown in Table~\ref{tab_bsqma111}. Interestingly, the power of the $\mathcal{H}$OP-tests is mostly highest at a delay of 4. With the exception of two cases for $(10, 10, 7)$, at a delay of 4, all $\mathcal{H}$OP-tests have a higher power than the parametric $\hat{\rho}(\1)$-test. Similarly to the results in Table~\ref{tab_sqma111}, the power of the $\mathcal{H}$OP-tests for grid size $(10, 10, 7)$ differ due to the resulting variation in the Hilbert curves. 

\smallskip
Finally, to substitute the (2D) AR-like DGPs \eqref{SAR11} and \eqref{simultAR1}, we simulate 3D Gaussian random fields (GRF) in the grid points $\mathcal{T}=\{0,\ldots,n_1\}\times\cdots\times \{0,\ldots,n_3\}$ by using the efficient implementation \href{https://pieterjanrobbe.github.io/GaussianRandomFields.jl/stable/}{\nolinkurl{GaussianRandomFields.jl}} in \texttt{Julia} by \citet{robbe23}. The GRF $(X_{\fti})_{\fti\in\mathcal{T}}$ is defined by the joint density
\ba
\label{GRF}
f(\fx) = \frac{1}{\sqrt{2\pi\,\det{\mC}}}\, \exp\Big(-\tfrac{1}{2}\,(\fx-\fmu)^\top\mC^{-1}(\fx-\fmu)\Big)
\ea
with covariance matrix~$\mC$, where $\fx$ abbreviates the possible outcome $(x_{\fti})_{\fti\in\mathcal{T}}$, and where~$\fmu$ comprises the corresponding means. Here, we consider a zero-mean GRF, where~$\mC$ is generated by the exponential covariance kernel $C(\ft_i, \ft_j)=\exp\big(-\textup{d}(\ft_i, \ft_j)/\lambda\big)$ with unit variance. We uniquely use the rescaled Euclidean distance 
$$
\textstyle
\textup{d}(\ft_i, \ft_j) = \left(\left(\frac{t_{i,1}-t_{j,1}}{15}\right)^2+\left(\frac{t_{i,2}-t_{j,2}}{15}\right)^2+\left(\frac{t_{i,3}-t_{j,3}}{7}\right)^2\right)^{1/2}
$$
together with the length scales $\lambda\in\{0.05,0.10\}$ (DGPs ``3D-3.1'' and ``3D-3.2'', respectively). Table~\ref{tab_grf} shows the corresponding rejection rates, and Table~\ref{tab_grf_ao} provides results for the same DGPs but with observations that are contaminated by AOs (10\,\% with value $+10$). As before, the $\mathcal{H}$OP-based $\hat{\uptau}$-test has always higher power than the entropy $\widehat{H}$-test. In addition, the power of $\hat{\uptau}$ at a delay of 1 is always greater than the power of $\hat{\rho}(\1)$. Although all approaches become less effective when dealing with AOs, the power of the parametric $\hat{\rho}(\1)$-test decreases vastly more than that of the nonparametric $\mathcal{H}$OP-tests. Similarly to before, we find substantial deviations in the power due to different Hilbert curves only for the particular grid size $(10,10,7)$.

\bigskip
Our power analyses for 3D scenarios mostly mirror the findings for the 2D scenarios, namely that our novel $\mathcal{H}$OP-tests, in particular $\hat{\uptau}$, are powerful and robust tools to detect spatial dependence in a variety of settings. In addition, the efficiency of the $\mathcal{H}$OP-tests often increases again at higher delays. This result highlights once more the possible benefit of an approach that can aggregate information across multiple delays, pointing to a fruitful endeavor for future research. 
While it is still true that the actual choice of the Hilbert curve is generally without great effect, for a grid of dimension $(10,10,7)$, we observed notable differences in the power of the same $\mathcal{H}$OP approaches, which stem from rather different Hilbert curves by the approaches of \href{https://github.com/jakubcerveny/gilbert}{\nolinkurl{gilbert}} and \citet{rong21}. 
%When confronted with the challenge of two Hilbert curves exhibiting distinct spatial paths, it could be more advantageous to harmonize both curves, or to investigate alternative algorithms. 
Hence, when confronted with a Hilbert curve that has an unusual appearance, it seems advantageous to investigate alternative algorithms in order to obtain a more ``harmonic'' Hilbert curve. This opens up further avenues for future research.

%Possible solution could be the Julia package \href{https://pieterjanrobbe.github.io/GaussianRandomFields.jl/stable/}{\nolinkurl{GaussianRandomFields}} of \citet{robbe23}, 
% or the Julia package \href{https://github.com/luraess/ParallelRandomFields.jl}{\nolinkurl{ParallelRandomFields}} of \citet{raess19}, 
%which offers, among others, Gaussian random fields in 2D and 3D with exponential autocorrelation function. 
%%Julia package \href{https://pieterjanrobbe.github.io/GaussianRandomFields.jl/stable/}{\nolinkurl{GaussianRandomFields}}

\section{Data Applications}
\label{Data Applications}

\subsection{Analysis of Art Paintings}
\label{Analysis of Art Paintings}
\begin{figure}[H]
\centering\footnotesize
\begin{tabular}{l@{\qquad}l}
\includegraphics[viewport=0 0 260 260, clip=, scale=0.65]{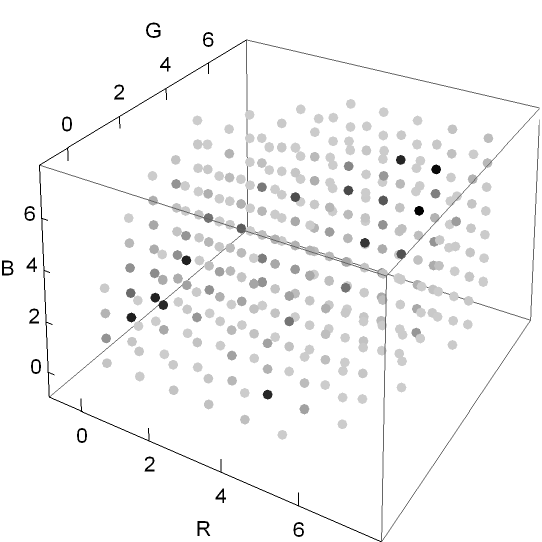}
&
\includegraphics[viewport=0 0 260 260, clip=, scale=0.65]{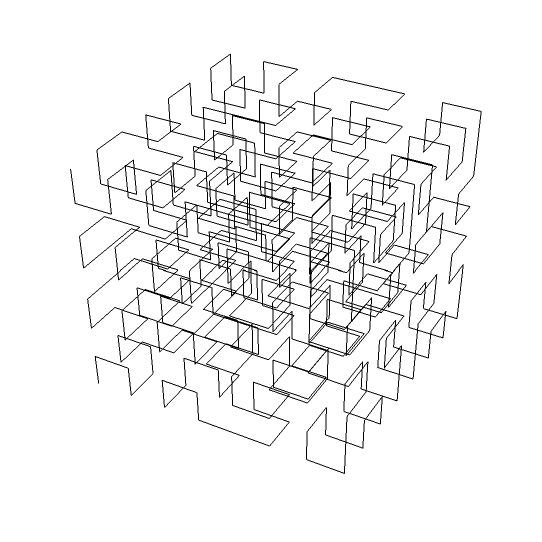}
\\[-3ex]
(a) & (b) \\[3ex]
\multicolumn{2}{l}{(c)\hspace{-3ex}%
\includegraphics[viewport=0 45 692 230, clip=, scale=0.5]{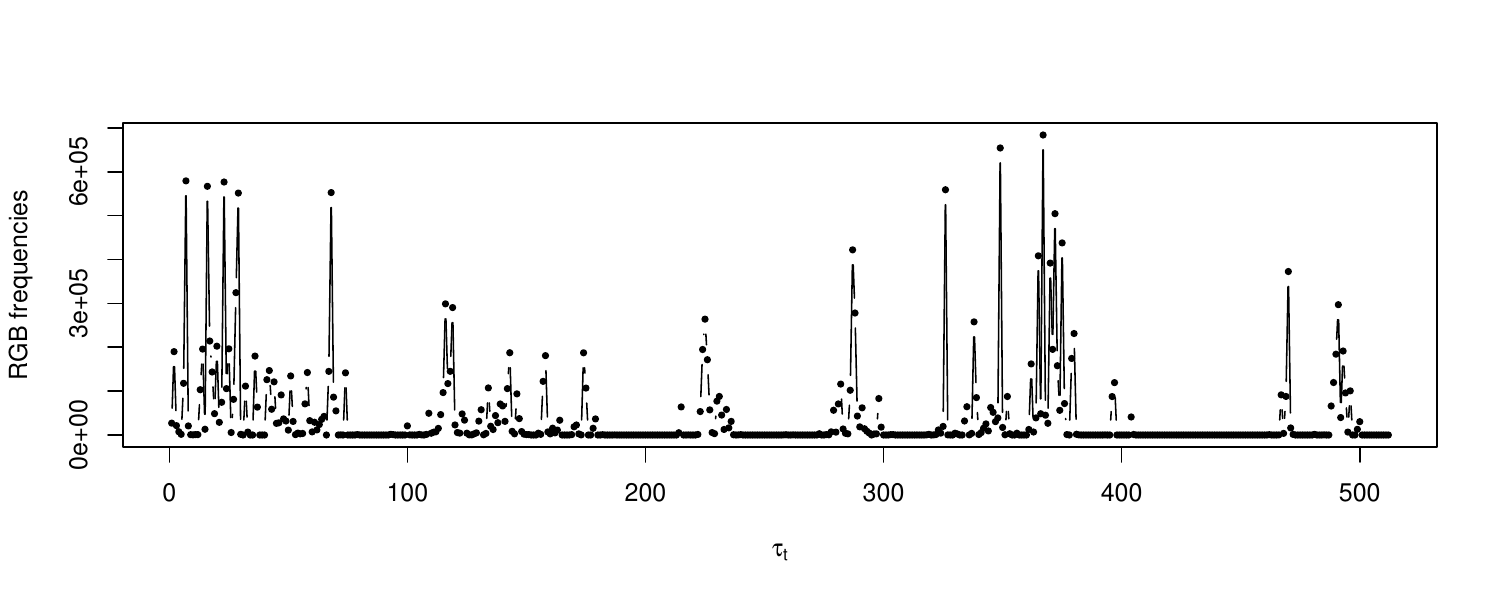}\!$\ftau_t$}
\end{tabular}
\caption{(a) RGB frequency data from Section~\ref{Analysis of Art Paintings}, where increasing darkness corresponds to increasing frequency (dots for zero frequency omitted). A corresponding 3D Hilbert curve is shown in (b), and the resulting time series plot in (c).}
\label{fig3Dcurve}
\end{figure}

As our first illustration, the proposed Hilbert-curve approach from Section~\ref{The Novel Hilbert-Curve Approach} is applied to a simple 3D data example. Inspired by the data application of \citet{sigaki18}, where art paintings are analyzed by SOPs, we create a 3D data cube from the work ``Blue Horse~I'' by Franz Marc, where a digital version with resolution $3881 \times 5180$ pixels is offered by the visual art encyclopedia \href{https://www.wikiart.org/en/franz-marc/blue-horse-i-1911}{\nolinkurl{WikiArt.org}} (accessed on May 28, 2025). The original file uses an 8-bit RGB color model, \ie each pixel corresponds to a vector from $\{0, \ldots, 2^8-1\}^3$, where the three components express the respective color depth of the three additive primary colors Red, Green, and Blue. We now created a 3D data cube (integer-valued) by computing the absolute frequency~$x_{\fti}$ for each possible RGB vector~$\ft$. In order to serve as an illustrative example, we reduced the color depth to 3~bit, \ie the final RGB frequency data are given by the cube $(x_{\fti})_{\fti\in\mathcal{T}}$ with $\mathcal{T}=\{0,\ldots,7\}^3$ and $|\mathcal{T}|=512$. These data are plotted as gray dots in Figure~\ref{fig3Dcurve}\,(a), where increasing darkness expresses increasing frequency. For the sake of readability, we omitted plotting the white dots expressing zero frequency. To illustrate the wide availability of Hilbert curves, this time, we used Wolfram Mathematica for generating the 3D Hilbert curve~$\mathcal{H}$, see the plot in part~(b), but we also cross-checked our final test decisions by the (somewhat different) Hilbert curves according to \href{https://github.com/jakubcerveny/gilbert}{\nolinkurl{gilbert}} and \citet{rong21}, recall the first column in Figure~\ref{figHilbert3D}.

\smallskip
According to the Hilbert-curve approach proposed in Section~\ref{The Novel Hilbert-Curve Approach}, we read the RGB frequencies~$x_{\fti}$ along the Hilbert curve~$\mathcal{H}$, which leads to the sequence $(x_{\ftaui_t})_{t=1,\ldots,512}$. The corresponding time series plot is shown in Figure~\ref{fig3Dcurve}\,(c), where most frequency values are quite low (including 209~zeros). However, the long phases of low frequencies are interrupted by clusters of large up to extreme frequencies, clearly indicating the presence of positive dependence in combination with heteroskedasticity. At this point, we recall that the extreme frequencies do not affect the subsequent $\mathcal{H}$OP-statistics, because these are robust against outliers due to being based on ranks. It should also be noted that the RGB frequencies are integer-valued and thus discrete. Therefore, we follow the approach sketched in Remark~\ref{bemTies} and first add uniform noise to $(x_{\ftaui_t})$ before computing the $\mathcal{H}$OP-statistics, in order to remove ties from the data (ties virtually happen between zeros only).

\begin{table}[t]
\centering
\caption{RGB frequency data from Section~\ref{Analysis of Art Paintings}: $\mathcal{H}$OP-statistics for delays $d=1,\ldots,5$ with corresponding P-values in parentheses.}
\label{tabRGB}

\smallskip
\begin{tabular}{c@{\qquad}c@{\qquad}c@{\qquad}c@{\qquad}c@{\qquad}c@{\qquad}c}
\toprule
$\widehat{H}^{(d)}$ & $\widehat{H}_{\textup{ex}}^{(d)}$ & $\widehat{\Delta}_2^{(d)}$ & $\hat{\upbeta}^{(d)}$ & $\hat{\uptau}^{(d)}$ & $\hat{\upgamma}^{(d)}$ & $\hat{\updelta}^{(d)}$ \\
\midrule
$d=1$ & \multicolumn{6}{l}{$\hat{\fp}^{(d)}\approx (0.208, 0.133, 0.143, 0.155, 0.143, 0.218)^\top$} \\[1ex]
1.772 & 0.908 & 0.007 & -0.010 & 0.092 & 0.022 & -0.022 \\
\footnotesize (0.001) & \footnotesize (0.001) & \footnotesize (0.001) & \footnotesize (0.701) & \footnotesize (0.000) & \footnotesize (0.441) & \footnotesize (0.551) \\
\midrule
$d=2$ & \multicolumn{6}{l}{$\hat{\fp}^{(d)}\approx (0.150, 0.167, 0.181, 0.169, 0.156, 0.177)^\top$} \\[1ex]
1.790 & 0.911 & 0.001 & -0.028 & -0.007 & 0.028 & 0.024 \\
\footnotesize (0.568) & \footnotesize (0.572) & \footnotesize (0.571) & \footnotesize (0.282) & \footnotesize (0.726) & \footnotesize (0.326) & \footnotesize (0.514) \\
\midrule
$d=3$ & \multicolumn{6}{l}{$\hat{\fp}^{(d)}\approx (0.188, 0.160, 0.146, 0.156, 0.168, 0.182)^\top$} \\[1ex]
1.788 & 0.911 & 0.001 & 0.006 & 0.036 & -0.026 & -0.018 \\
\footnotesize (0.328) & \footnotesize (0.326) & \footnotesize (0.327) & \footnotesize (0.817) & \footnotesize (0.053) & \footnotesize (0.361) & \footnotesize (0.624) \\
\midrule
$d=4$ & \multicolumn{6}{l}{$\hat{\fp}^{(d)}\approx (0.198, 0.155, 0.139, 0.145, 0.161, 0.202)^\top$} \\[1ex]
1.781 & 0.909 & 0.004 & -0.004 & 0.067 & -0.032 & -0.012 \\
\footnotesize (0.024) & \footnotesize (0.022) & \footnotesize (0.022) & \footnotesize (0.877) & \footnotesize (0.000) & \footnotesize (0.260) & \footnotesize (0.743) \\
\midrule
$d=5$ & \multicolumn{6}{l}{$\hat{\fp}^{(d)}\approx (0.201, 0.141, 0.153, 0.137, 0.127, 0.239)^\top$} \\[1ex]
1.764 & 0.906 & 0.010 & -0.038 & 0.107 & 0.022 & 0.030 \\
\footnotesize (0.000) & \footnotesize (0.000) & \footnotesize (0.000) & \footnotesize (0.142) & \footnotesize (0.000) & \footnotesize (0.438) & \footnotesize (0.412) \\
\bottomrule
\end{tabular}
\end{table}

\smallskip
The computed $\mathcal{H}$OP-statistics are summarized in Table~\ref{tabRGB} for the delays $d=1,\ldots,5$, where the corresponding P-values are shown in parentheses. We recognize a quite sophisticated dependence structure with significant values for delays $d=1, 4, 5$. More precisely, for these delays, only the three entropy-like statistics \eqref{entropies} as well as the $\hat{\uptau}$-statistic from \eqref{LinStat3} are significant on the 5\,\%-level. This result can be well explained by our performance analyses in Section~\ref{Performance Analyses}, and it also fits to the power analyses in \citet{weiss22}. 
%The three entropy-like tests from Proposition~\ref{propEntropies} were shown to be ``universally applicable'', \ie with a reasonable power in most dependence scenarios ($\approx$ omnibus property). Among the tests using a linear statistic, see Proposition~\ref{propLinearStatistics}, the $\hat{\uptau}$-test showed an outstanding performance, whereas the $\hat{\upbeta}$- and $\hat{\updelta}$-test were powerful only for a few DGPs, and the $\hat{\upgamma}$-test was never powerful at all. 
In fact, if looking at the $\mathcal{H}$OP-frequencies~$\hat{\fp}{}^{(d)}$, we recognize that they are close to uniformity for $d=2$ and~3 (hence no significant dependence), while the deviations from uniformity for $d=1, 4, 5$ manifest themselves by increased values for~$\hat{p}_1{}^{(d)}$ and~$\hat{p}_6{}^{(d)}$. These increases are well recognized by the $\hat{\uptau}$-statistic according to \eqref{LinStat3}, which compares the sum $\hat{p}_1{}^{(d)}+\hat{p}_6{}^{(d)}$ to the null value of~$1/3$, and certainly by the entropies \eqref{entropies}, which have a kind of omnibus property concerning deviations from the discrete uniform distribution. But the $\hat{\upbeta}$-statistic, for example, which considers the difference $\hat{p}_1{}^{(d)}-\hat{p}_6{}^{(d)}$, does not recognize a violation of the null as long as~$\hat{p}_1{}^{(d)}$ and~$\hat{p}_6{}^{(d)}$ are increased simultaneously. Recall that the frequencies~$\hat{p}_1{}^{(d)}$ and~$\hat{p}_6{}^{(d)}$ refer to the strictly monotone $\mathcal{H}$OPs according to \eqref{OrdPatt3}, which are increased if the Hilbert curve~$\mathcal{H}$ enters or leaves a cluster, respectively, with increased RGB frequencies. Altogether, it gets clear that Franz Marc did not choose the colors randomly for his work ``Blue Horse~I'', but mainly from selected regions (clusters) in the color space.

\begin{figure}[t]
\centering\footnotesize
\includegraphics[viewport=35 45 340 240, clip=, scale=0.5]{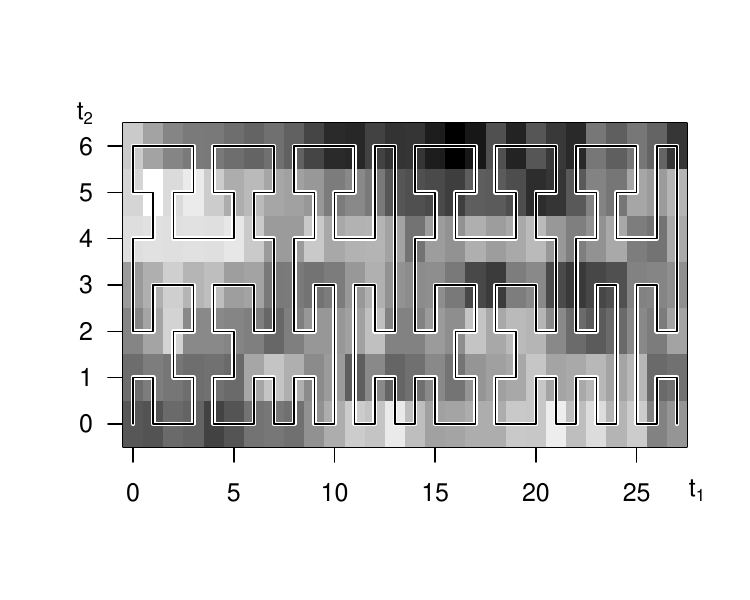}
\qquad
\includegraphics[viewport=0 45 340 240, clip=, scale=0.5]{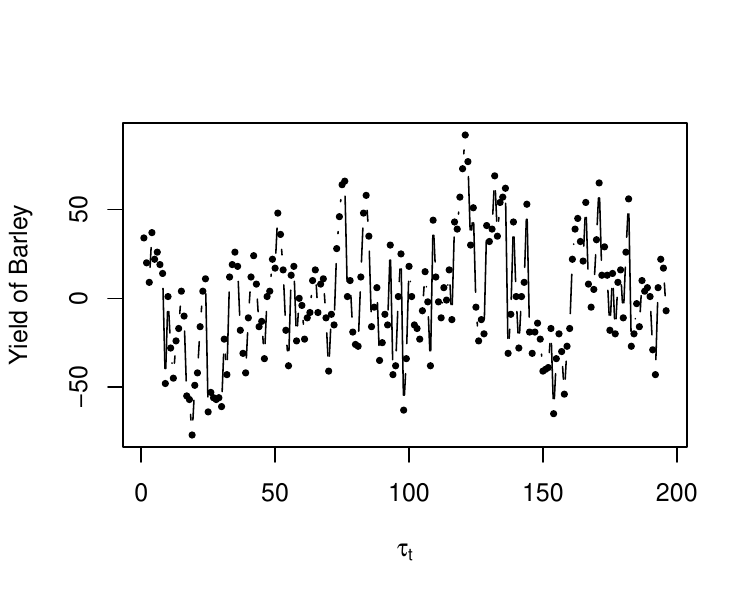}$\ftau_t$
\\[2ex]
\includegraphics[viewport=35 45 340 240, clip=, scale=0.5]{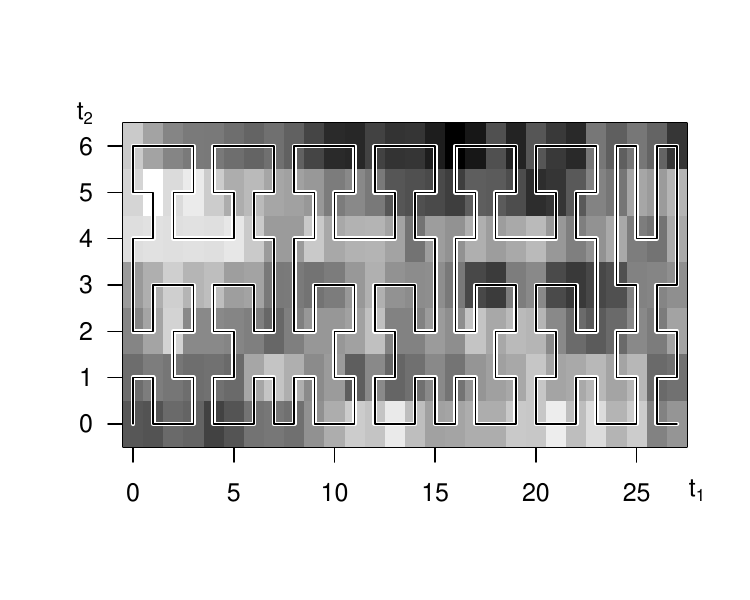}
\qquad
\includegraphics[viewport=0 45 340 240, clip=, scale=0.5]{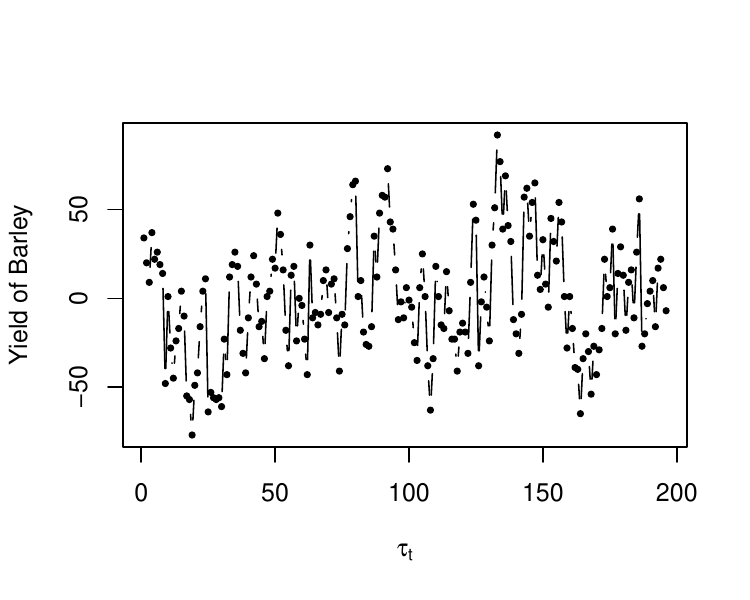}$\ftau_t$
\caption{Barley data from Section~\ref{Yield of Barley}: plot of data with superimposed Hilbert curve (left; increasing darkness for increasing yield) and resulting time series plot (right). Hilbert curve in first row computed by \href{https://github.com/jakubcerveny/gilbert}{\nolinkurl{gilbert}}, in second row by \citet{rong21}.}
\label{figBarley}
\end{figure}

\subsection{Yield of Barley}
\label{Yield of Barley}
As our 2D data application, we consider an example that was already successfully analyzed by SOP-based dependence tests, namely the yield of barley in an agricultural uniformity trial experiment, $(x_{\fti})$ with $t_1=0,\ldots,27$ and $t_2=0,\ldots,6$. The original data source is \citet{kempton81}, and the SOP-analyses can be found in \citet{WeissKim2024,WeissKim2025}, where significant spatial dependence was recognized as caused by the actual ``sowing, harvesting and all intermediate farming practices''. Plots of the data are shown in the left part of Figure~\ref{figBarley} (with increasing darkness expressing increasing yield), where two kinds of Hilbert curve are superimposed: the one by \href{https://github.com/jakubcerveny/gilbert}{\nolinkurl{gilbert}} in the first row, and the one by \citet{rong21} in the second row. The resulting sequences $(x_{\ftaui_t})_{t=1,\ldots,196}$ are shown in the respective right part of Figure~\ref{figBarley}.

\begin{table}[th!]
\centering
\caption{Barley data from Section~\ref{Yield of Barley}: $\mathcal{H}$OP-statistics for delays $d=1,2,3$ with corresponding P-values in parentheses. Upper table based on Hilbert curve by \href{https://github.com/jakubcerveny/gilbert}{\nolinkurl{gilbert}}, lower table by \citet{rong21}.}
\label{tabBarley}

\smallskip
\begin{tabular}{c@{\qquad}c@{\qquad}c@{\qquad}c@{\qquad}c@{\qquad}c@{\qquad}c}
\toprule
$\widehat{H}^{(d)}$ & $\widehat{H}_{\textup{ex}}^{(d)}$ & $\widehat{\Delta}_2^{(d)}$ & $\hat{\upbeta}^{(d)}$ & $\hat{\uptau}^{(d)}$ & $\hat{\upgamma}^{(d)}$ & $\hat{\updelta}^{(d)}$ \\
\midrule
$d=1$ & \multicolumn{6}{l}{$\hat{\fp}^{(d)}\approx (0.258, 0.124, 0.149, 0.149, 0.124, 0.196)^\top$}    \\[1ex]
1.754 & 0.903 & 0.013 & 0.062 & 0.120 & 0.052 & 0.000 \\
\footnotesize (0.007) & \footnotesize (0.005) & \footnotesize (0.005) & \footnotesize (0.136) & \footnotesize (0.000) & \footnotesize (0.256) & \footnotesize (1.000) \\
\midrule
$d=2$ & \multicolumn{6}{l}{$\hat{\fp}^{(d)}\approx (0.229, 0.135, 0.125, 0.167, 0.177, 0.167)^\top$} \\[1ex]
1.772 & 0.908 & 0.007 & 0.063 & 0.063 & -0.021 & -0.083 \\
\footnotesize (0.081) & \footnotesize (0.074) & \footnotesize (0.076) & \footnotesize (0.134) & \footnotesize (0.040) & \footnotesize (0.648) & \footnotesize (0.157) \\
\midrule
$d=3$ & \multicolumn{6}{l}{$\hat{\fp}^{(d)}\approx (0.189, 0.137, 0.142, 0.179, 0.168, 0.184)^\top$} \\[1ex]
1.784 & 0.910 & 0.002 & 0.005 & 0.040 & 0.016 & -0.068 \\
\footnotesize (0.457) & \footnotesize (0.469) & \footnotesize (0.467) & \footnotesize (0.900) & \footnotesize (0.187) & \footnotesize (0.731) & \footnotesize (0.248) \\
%\midrule
%$d=4$ & \multicolumn{6}{l}{$\hat{\fp}^{(d)}\approx (0.144, 0.160, 0.160, 0.176, 0.186, 0.176)^\top$} \\[1ex]
%1.788 & 0.911 & 0.001 & -0.032 & -0.014 & -0.011 & -0.043 \\
%\footnotesize (0.775) & \footnotesize (0.778) & \footnotesize (0.778) & \footnotesize (0.448) & \footnotesize (0.645) & \footnotesize (0.818) & \footnotesize (0.475) \\
%\midrule
%$d=5$ & \multicolumn{6}{l}{$\hat{\fp}^{(d)}\approx (0.194, 0.151, 0.151, 0.172, 0.188, 0.145)^\top$} \\[1ex]
%1.785 & 0.910 & 0.002 & 0.048 & 0.005 & -0.016 & -0.059 \\
%\footnotesize (0.537) & \footnotesize (0.533) & \footnotesize (0.534) & \footnotesize (0.253) & \footnotesize (0.862) & \footnotesize (0.728) & \footnotesize (0.323) \\
\bottomrule
\end{tabular}

\vspace{1ex}
\begin{tabular}{c@{\qquad}c@{\qquad}c@{\qquad}c@{\qquad}c@{\qquad}c@{\qquad}c}
\toprule
$\widehat{H}^{(d)}$ & $\widehat{H}_{\textup{ex}}^{(d)}$ & $\widehat{\Delta}_2^{(d)}$ & $\hat{\upbeta}^{(d)}$ & $\hat{\uptau}^{(d)}$ & $\hat{\upgamma}^{(d)}$ & $\hat{\updelta}^{(d)}$ \\
\midrule
$d=1$ & \multicolumn{6}{l}{$\hat{\fp}^{(d)}\approx (0.216, 0.139, 0.165, 0.124, 0.098, 0.258)^\top$} \\[1ex]
1.739 & 0.901 & 0.018 & -0.041 & 0.141 & 0.052 & 0.082 \\
\footnotesize (0.001) & \footnotesize (0.001) & \footnotesize (0.001) & \footnotesize (0.320) & \footnotesize (0.000) & \footnotesize (0.256) & \footnotesize (0.159) \\
\midrule
$d=2$ & \multicolumn{6}{l}{$\hat{\fp}^{(d)}\approx (0.224, 0.135, 0.125, 0.167, 0.172, 0.177)^\top$} \\[1ex]
1.774 & 0.908 & 0.006 & 0.047 & 0.068 & -0.016 & -0.078 \\
\footnotesize (0.101) & \footnotesize (0.095) & \footnotesize (0.097) & \footnotesize (0.261) & \footnotesize (0.026) & \footnotesize (0.732) & \footnotesize (0.185) \\
\midrule
$d=3$ & \multicolumn{6}{l}{$\hat{\fp}^{(d)}\approx (0.211, 0.142, 0.132, 0.174, 0.179, 0.163)^\top$} \\[1ex]
1.780 & 0.909 & 0.004 & 0.047 & 0.040 & -0.016 & -0.079 \\
\footnotesize (0.251) & \footnotesize (0.247) & \footnotesize (0.248) & \footnotesize (0.258) & \footnotesize (0.187) & \footnotesize (0.731) & \footnotesize (0.183) \\
%\midrule
%$d=4$ & \multicolumn{6}{l}{$\hat{\fp}^{(d)}\approx (0.144, 0.191, 0.165, 0.144, 0.176, 0.181)^\top$} \\[1ex]
%1.786 & 0.910 & 0.002 & -0.037 & -0.009 & -0.059 & 0.037 \\
%\footnotesize (0.577) & \footnotesize (0.581) & \footnotesize (0.580) & \footnotesize (0.377) & \footnotesize (0.773) & \footnotesize (0.205) & \footnotesize (0.532) \\
%\midrule
%$d=5$ & \multicolumn{6}{l}{$\hat{\fp}^{(d)}\approx (0.118, 0.183, 0.194, 0.167, 0.172, 0.167)^\top$} \\[1ex]
%1.781 & 0.910 & 0.003 & -0.048 & -0.048 & 0.005 & 0.038 \\
%\footnotesize (0.303) & \footnotesize (0.337) & \footnotesize (0.332) & \footnotesize (0.253) & \footnotesize (0.118) & \footnotesize (0.908) & \footnotesize (0.530) \\
\bottomrule
\end{tabular}
\end{table}

\smallskip
In Table~\ref{tabBarley}, we applied the $\mathcal{H}$OP-tests for delays $d=1,2,3$ to both kinds of $\mathcal{H}$OP-sequence. It can be recognized that for both Hilbert curves, we get a clear rejection of the null of spatial independence on the 5\,\%-level. This test decision is caused by all entropy-like $\mathcal{H}$OP-tests with delay $d=1$, and by the $\hat{\uptau}$-test with delays $d=1$ and~$2$, in agreement to the power analyses in Section~\ref{Performance Analyses}. 
Furthermore, Table~\ref{tabBarley} also illustrates one of the main findings from our performance analyses in Section~\ref{Performance Analyses}: while there are different ways of computing Hilbert curves, and while these affect the actual values of the $\mathcal{H}$OP-statistics and the respective power performance, these differences are practically negligible as we end up with the same test decisions. 
Altogether, our test decisions confirm the findings of \citet{WeissKim2024,WeissKim2025}, namely pronounced spatial in the barley data, and thus demonstrate that $\mathcal{H}$OP-tests might be used as a suitable alternative to the existing SOP-tests. In particular, recall that for 3D or higher-dimensional data (like in the example of Section~\ref{Analysis of Art Paintings}), there do not exist any SOP-tests, \ie in these cases, the $\mathcal{H}$OP-tests are even indispensable.

\section{Conclusions}
\label{Conclusions}
We have introduced a flexible nonparametric framework for testing spatial dependence in 2D and 3D random fields by transforming spatial grid data into a one-dimensional sequence via space-filling Hilbert curves. This transformation enables the use of established OP-based tests for serial dependence in a spatial context, providing a simple and broadly applicable alternative to existing methods. Our simulation study shows that, compared to the classical SACF, the nonparametric approach, in particular~$\widehat{H}$ and~$\hat{\uptau}$, performs particularly well for nonlinear DGPs or those containing additive outliers. By contrast, in linear and parametric settings, the classical ACF generally outperforms the nonparametric approach. While SOP-based tests remain more powerful in purely 2D settings, the proposed $\mathcal{H}$OP-approach using $\hat{\uptau}$ is still competitive and offers the advantage of general applicability in higher dimensions. With one exception, the results were largely insensitive to the specific method used to compute the Hilbert curves, underscoring the robustness of the proposed approach. 

\smallskip
There are several promising directions for future research. One possibility is to adapt the Hilbert-curve approach to the sequential monitoring of spatial dependence by using control charts, similar to \citet{WeissTestik2023,adaemmer25}. Another possibility is to extend the method to discrete-valued data (recall Remark~\ref{bemTies}) following the approach of \citet{WeissSchnurr2024}. Since spatial dependence can induce serial dependence on delays $d>1$ in Hilbert-transformed time series, novel tests could be developed that aggregate information over multiple delays into a single statistic, analogous to the classic Box--Pierce test in time series analysis. This approach could allow researchers to more generally account for dependence across multiple delays, thereby reducing the need to select an ``optimal'' delay in advance.
%Finally, exploring applications of our $\mathcal{H}$OP-tests to 3D grid data obtained from computer-aided tomography—common in biological and medical contexts—offers a practical and impactful research direction, as exemplified by \citet{orland16,marwan07,kochetkova25}.

\subsubsection*{Acknowledgements}
%The authors thank the editor, the associate editor, and the referees for their useful comments on an earlier draft of this article.
The authors are grateful to Professor Aurelio F.\ Bariviera (Universitat Rovira i Virgili, Spain) for helpful advice on Hilbert curves.
The authors are also indebted to Professor Annika Betken (University of Twente) and Professor Alexander Schnurr (University of Siegen) for inviting them to participate in the COBAP workshop 2025 at the University of Twente, which served as the inspiration for this research.

%\subsubsection*{Data Availability}

%\spacingset{1}

\clearpage

\appendix
\small
\numberwithin{equation}{section}
\numberwithin{table}{section}

\section{Tabulated Simulation Results}
\label{Tabulated Simulation Results}

%Two columns like in Chaos' journal layout

\begin{table}[hb!]
\centering
\begin{multicols}{2}
\caption{Simulated power for SAR$(1,1)$-DGPs with sample size $(n_1,n_2)$, where Hilbert curve by \href{https://github.com/jakubcerveny/gilbert}{\nolinkurl{gilbert}} or \citet{rong21}, respectively. Italic numbers taken from \citet{WeissKim2024}.}
\label{tab_sar11}

\smallskip
\resizebox{.9\linewidth}{!}{
\begin{tabular}{lc@{\qquad}c@{\qquad}c@{\qquad}c}
\toprule
Delay & \multicolumn{2}{c}{``gilbert''} & \multicolumn{2}{c}{``Rong''} \\
$d$ & $\widehat{H}$ & $\hat{\uptau}$ & $\widehat{H}$ & $\hat{\uptau}$ \\
\midrule
2D-1.1 & \multicolumn{4}{l}{$(10, 10)$,\quad $\widetilde{\tau}$: {\it 0.098},\quad $\hat{\rho}(\1)$: {\it 0.710}} \\
\midrule
1 & 0.068 & 0.118 & 0.066 & 0.114 \\
2 & 0.057 & 0.083 & 0.059 & 0.086 \\
3 & 0.060 & 0.065 & 0.062 & 0.069 \\
4 & 0.058 & 0.059 & 0.060 & 0.060 \\
%5 & 0.060 & 0.069 & 0.060 & 0.072 \\
\midrule
2D-1.1 & \multicolumn{4}{l}{$(15, 15)$,\quad $\widetilde{\tau}$: {\it 0.167},\quad $\hat{\rho}(\1)$: {\it 0.982}} \\
\midrule
1 & 0.082 & 0.169 & 0.082 & 0.169 \\
2 & 0.067 & 0.091 & 0.065 & 0.092 \\
3 & 0.063 & 0.092 & 0.065 & 0.093 \\
4 & 0.057 & 0.073 & 0.058 & 0.071 \\
%5 & 0.059 & 0.082 & 0.057 & 0.082 \\
\midrule
2D-1.1 & \multicolumn{4}{l}{$(20, 20)$,\quad $\widetilde{\tau}$: {\it 0.318},\quad $\hat{\rho}(\1)$: {\it 1.000}} \\
\midrule
1 & 0.121 & 0.276 & 0.120 & 0.269 \\
2 & 0.066 & 0.114 & 0.070 & 0.128 \\
3 & 0.071 & 0.113 & 0.065 & 0.098 \\
4 & 0.060 & 0.080 & 0.058 & 0.074 \\
%5 & 0.060 & 0.085 & 0.058 & 0.081 \\
\midrule
2D-1.1 & \multicolumn{4}{l}{$(40, 25)$,\quad $\widetilde{\tau}$: {\it 0.641},\quad $\hat{\rho}(\1)$: {\it 1.000}} \\
\midrule
1 & 0.249 & 0.523 & 0.223 & 0.481 \\
2 & 0.100 & 0.213 & 0.115 & 0.260 \\
3 & 0.107 & 0.218 & 0.106 & 0.215 \\
4 & 0.065 & 0.104 & 0.071 & 0.113 \\
%5 & 0.064 & 0.100 & 0.072 & 0.123 \\
\midrule
2D-1.2 & \multicolumn{4}{l}{$(10, 10)$,\quad $\widetilde{\tau}$: {\it 0.504},\quad $\hat{\rho}(\1)$: {\it 0.866}} \\
\midrule
1 & 0.147 & 0.349 & 0.142 & 0.337 \\
2 & 0.066 & 0.115 & 0.065 & 0.112 \\
3 & 0.077 & 0.106 & 0.084 & 0.126 \\
4 & 0.071 & 0.089 & 0.078 & 0.109 \\
%5 & 0.078 & 0.101 & 0.083 & 0.117 \\
\midrule
2D-1.2 & \multicolumn{4}{l}{$(15, 15)$,\quad $\widetilde{\tau}$: {\it 0.844},\quad $\hat{\rho}(\1)$: {\it 0.997}} \\
\midrule
1 & 0.268 & 0.574 & 0.267 & 0.575 \\
2 & 0.078 & 0.124 & 0.079 & 0.125 \\
3 & 0.105 & 0.205 & 0.104 & 0.203 \\
4 & 0.093 & 0.173 & 0.092 & 0.173 \\
%5 & 0.085 & 0.156 & 0.085 & 0.158 \\
\midrule
2D-1.2 & \multicolumn{4}{l}{$(20, 20)$,\quad $\widetilde{\tau}$: {\it 0.985},\quad $\hat{\rho}(\1)$: {\it 1.000}} \\
\midrule
1 & 0.525 & 0.828 & 0.521 & 0.825 \\
2 & 0.092 & 0.199 & 0.096 & 0.206 \\
3 & 0.127 & 0.259 & 0.121 & 0.245 \\
4 & 0.088 & 0.162 & 0.098 & 0.189 \\
%5 & 0.096 & 0.185 & 0.092 & 0.179 \\
\midrule
2D-1.2 & \multicolumn{4}{l}{$(40, 25)$,\quad $\widetilde{\tau}$: {\it 1.000},\quad $\hat{\rho}(\1)$: {\it 1.000}} \\
\midrule
1 & 0.920 & 0.992 & 0.913 & 0.991 \\
2 & 0.142 & 0.315 & 0.165 & 0.365 \\
3 & 0.335 & 0.616 & 0.334 & 0.609 \\
4 & 0.175 & 0.365 & 0.237 & 0.480 \\
%5 & 0.144 & 0.305 & 0.188 & 0.395 \\
\bottomrule
\end{tabular}}
%\end{table}

%\vfill\null
%\columnbreak

%\begin{table}{ht}
%\centering
\caption{Simulated power for SAR$(1,1)$-DGPs with sample size $(n_1,n_2)$ like in Table~\ref{tab_sar11}, but data contaminated by AOs. Italic numbers taken from \citet{WeissKim2024}.\newline\mbox{}}
\label{tab_sar11_ao}

\smallskip
\resizebox{.9\linewidth}{!}{
\begin{tabular}{lc@{\qquad}c@{\qquad}c@{\qquad}c}
\toprule
Delay & \multicolumn{2}{c}{``gilbert''} & \multicolumn{2}{c}{``Rong''} \\
$d$ & $\widehat{H}$ & $\hat{\uptau}$ & $\widehat{H}$ & $\hat{\uptau}$ \\
\midrule
2D-2.1 & \multicolumn{4}{l}{$(10, 10)$,\quad $\widetilde{\tau}$: {\it 0.071},\quad $\hat{\rho}(\1)$: {\it 0.036}} \\
\midrule
1 & 0.061 & 0.088 & 0.060 & 0.086 \\
2 & 0.056 & 0.074 & 0.056 & 0.078 \\
3 & 0.056 & 0.056 & 0.058 & 0.058 \\
4 & 0.057 & 0.053 & 0.056 & 0.054 \\
%5 & 0.058 & 0.064 & 0.059 & 0.065 \\
\midrule
2D-2.1 & \multicolumn{4}{l}{$(15, 15)$,\quad $\widetilde{\tau}$: {\it 0.106},\quad $\hat{\rho}(\1)$: {\it 0.059}} \\
\midrule
1 & 0.068 & 0.116 & 0.067 & 0.118 \\
2 & 0.059 & 0.070 & 0.059 & 0.069 \\
3 & 0.056 & 0.072 & 0.058 & 0.075 \\
4 & 0.055 & 0.062 & 0.055 & 0.062 \\
%5 & 0.056 & 0.072 & 0.056 & 0.072 \\
\midrule
2D-2.1 & \multicolumn{4}{l}{$(20, 20)$,\quad $\widetilde{\tau}$: {\it 0.184},\quad $\hat{\rho}(\1)$: {\it 0.091}} \\
\midrule
1 & 0.085 & 0.170 & 0.085 & 0.169 \\
2 & 0.060 & 0.088 & 0.058 & 0.094 \\
3 & 0.062 & 0.083 & 0.059 & 0.075 \\
4 & 0.056 & 0.068 & 0.054 & 0.064 \\
%5 & 0.055 & 0.069 & 0.055 & 0.069 \\
\midrule
2D-2.1 & \multicolumn{4}{l}{$(40, 25)$,\quad $\widetilde{\tau}$: {\it 0.384},\quad $\hat{\rho}(\1)$: {\it 0.180}} \\
\midrule
1 & 0.140 & 0.319 & 0.130 & 0.292 \\
2 & 0.077 & 0.141 & 0.082 & 0.163 \\
3 & 0.076 & 0.140 & 0.077 & 0.138 \\
4 & 0.058 & 0.078 & 0.061 & 0.084 \\
%5 & 0.057 & 0.079 & 0.060 & 0.091 \\
\midrule
2D-2.2 & \multicolumn{4}{l}{$(10, 10)$,\quad $\widetilde{\tau}$: {\it 0.289},\quad $\hat{\rho}(\1)$: {\it 0.058}} \\
\midrule
1 & 0.098 & 0.215 & 0.097 & 0.209 \\
2 & 0.059 & 0.092 & 0.060 & 0.091 \\
3 & 0.067 & 0.080 & 0.068 & 0.091 \\
4 & 0.064 & 0.070 & 0.068 & 0.080 \\
%5 & 0.067 & 0.079 & 0.071 & 0.090 \\
\midrule
2D-2.2 & \multicolumn{4}{l}{$(15, 15)$,\quad $\widetilde{\tau}$: {\it 0.558},\quad $\hat{\rho}(\1)$: {\it 0.117}} \\
\midrule
1 & 0.152 & 0.356 & 0.151 & 0.358 \\
2 & 0.064 & 0.088 & 0.066 & 0.088 \\
3 & 0.078 & 0.135 & 0.076 & 0.132 \\
4 & 0.073 & 0.116 & 0.072 & 0.118 \\
%5 & 0.070 & 0.115 & 0.069 & 0.110 \\
\midrule
2D-2.2 & \multicolumn{4}{l}{$(20, 20)$,\quad $\widetilde{\tau}$: {\it 0.832},\quad $\hat{\rho}(\1)$: {\it 0.199}} \\
\midrule
1 & 0.277 & 0.573 & 0.278 & 0.570 \\
2 & 0.072 & 0.134 & 0.072 & 0.137 \\
3 & 0.088 & 0.160 & 0.086 & 0.154 \\
4 & 0.070 & 0.111 & 0.075 & 0.124 \\
%5 & 0.073 & 0.123 & 0.072 & 0.122 \\
\midrule
2D-2.2 & \multicolumn{4}{l}{$(40, 25)$,\quad $\widetilde{\tau}$: {\it 0.997},\quad $\hat{\rho}(\1)$: {\it 0.481}} \\
\midrule
1 & 0.634 & 0.895 & 0.618 & 0.884 \\
2 & 0.094 & 0.199 & 0.105 & 0.226 \\
3 & 0.180 & 0.383 & 0.184 & 0.385 \\
4 & 0.108 & 0.225 & 0.137 & 0.294 \\
%5 & 0.096 & 0.192 & 0.116 & 0.243 \\
\bottomrule
\end{tabular}}
\end{multicols}
\end{table}

\begin{table}[ht!]
\centering
\begin{multicols}{2}
\caption{Simulated power for multilateral SAR$(1)$-DGPs with sample size $(n_1,n_2)$, where Hilbert curve by \href{https://github.com/jakubcerveny/gilbert}{\nolinkurl{gilbert}} or \citet{rong21}, respectively. Italic numbers taken from \citet{WeissKim2024}.}
\label{tab_sar1}

\smallskip
\resizebox{.9\linewidth}{!}{
\begin{tabular}{lc@{\qquad}c@{\qquad}c@{\qquad}c}
\toprule
Delay & \multicolumn{2}{c}{``gilbert''} & \multicolumn{2}{c}{``Rong''} \\
$d$ & $\widehat{H}$ & $\hat{\uptau}$ & $\widehat{H}$ & $\hat{\uptau}$ \\
\midrule
2D-3.1 & \multicolumn{4}{l}{$(10, 10)$,\quad $\widetilde{\tau}$: {\it 0.247},\quad $\hat{\rho}(\1)$: {\it 0.056}} \\
\midrule
1 & 0.086 & 0.175 & 0.083 & 0.167 \\
2 & 0.059 & 0.062 & 0.044 & 0.075 \\
3 & 0.056 & 0.058 & 0.059 & 0.051 \\
4 & 0.052 & 0.050 & 0.065 & 0.050 \\
%5 & 0.062 & 0.065 & 0.064 & 0.058 \\
\midrule
2D-3.1 & \multicolumn{4}{l}{$(15, 15)$,\quad $\widetilde{\tau}$: {\it 0.480},\quad $\hat{\rho}(\1)$: {\it 0.122}} \\
\midrule
1 & 0.110 & 0.260 & 0.103 & 0.255 \\
2 & 0.038 & 0.051 & 0.042 & 0.059 \\
3 & 0.056 & 0.072 & 0.049 & 0.071 \\
4 & 0.048 & 0.056 & 0.060 & 0.055 \\
%5 & 0.056 & 0.068 & 0.057 & 0.073 \\
\midrule
2D-3.1 & \multicolumn{4}{l}{$(20, 20)$,\quad $\widetilde{\tau}$: {\it 0.780},\quad $\hat{\rho}(\1)$: {\it 0.213}} \\
\midrule
1 & 0.194 & 0.420 & 0.178 & 0.432 \\
2 & 0.062 & 0.066 & 0.041 & 0.061 \\
3 & 0.053 & 0.083 & 0.081 & 0.087 \\
4 & 0.047 & 0.056 & 0.051 & 0.054 \\
%5 & 0.061 & 0.061 & 0.049 & 0.082 \\
\midrule
2D-3.1 & \multicolumn{4}{l}{$(40, 25)$,\quad $\widetilde{\tau}$: {\it 0.992},\quad $\hat{\rho}(\1)$: {\it 0.493}} \\
\midrule
1 & 0.433 & 0.770 & 0.478 & 0.773 \\
2 & 0.052 & 0.068 & 0.069 & 0.089 \\
3 & 0.083 & 0.154 & 0.066 & 0.139 \\
4 & 0.052 & 0.068 & 0.050 & 0.069 \\
%5 & 0.059 & 0.068 & 0.063 & 0.088 \\
\midrule
2D-3.2 & \multicolumn{4}{l}{$(10, 10)$,\quad $\widetilde{\tau}$: {\it 0.252},\quad $\hat{\rho}(\1)$: {\it 0.065}} \\
\midrule
1 & 0.064 & 0.144 & 0.068 & 0.159 \\
2 & 0.043 & 0.069 & 0.056 & 0.075 \\
3 & 0.069 & 0.071 & 0.049 & 0.054 \\
4 & 0.053 & 0.058 & 0.059 & 0.060 \\
%5 & 0.053 & 0.077 & 0.067 & 0.071 \\
\midrule
2D-3.2 & \multicolumn{4}{l}{$(15, 15)$,\quad $\widetilde{\tau}$: {\it 0.472},\quad $\hat{\rho}(\1)$: {\it 0.136}} \\
\midrule
1 & 0.111 & 0.274 & 0.134 & 0.280 \\
2 & 0.042 & 0.034 & 0.051 & 0.051 \\
3 & 0.053 & 0.080 & 0.062 & 0.074 \\
4 & 0.053 & 0.054 & 0.059 & 0.068 \\
%5 & 0.049 & 0.074 & 0.059 & 0.079 \\
\midrule
2D-3.2 & \multicolumn{4}{l}{$(20, 20)$,\quad $\widetilde{\tau}$: {\it 0.774},\quad $\hat{\rho}(\1)$: {\it 0.231}} \\
\midrule
1 & 0.181 & 0.436 & 0.195 & 0.426 \\
2 & 0.044 & 0.054 & 0.052 & 0.051 \\
3 & 0.061 & 0.085 & 0.056 & 0.074 \\
4 & 0.053 & 0.062 & 0.044 & 0.053 \\
%5 & 0.054 & 0.065 & 0.045 & 0.071 \\
\midrule
2D-3.2 & \multicolumn{4}{l}{$(40, 25)$,\quad $\widetilde{\tau}$: {\it 0.991},\quad $\hat{\rho}(\1)$: {\it 0.536}} \\
\midrule
1 & 0.454 & 0.774 & 0.424 & 0.736 \\
2 & 0.044 & 0.071 & 0.044 & 0.078 \\
3 & 0.080 & 0.158 & 0.078 & 0.128 \\
4 & 0.051 & 0.051 & 0.056 & 0.079 \\
%5 & 0.060 & 0.070 & 0.063 & 0.097 \\
\bottomrule
\end{tabular}}
%\end{table}

%\vfill\null
%\columnbreak

%\begin{table}{ht}
%\centering
\caption{Simulated power for multilateral SAR$(1)$-DGPs with sample size $(n_1,n_2)$ like in Table~\ref{tab_sar1}, but data contaminated by AOs. Italic numbers taken from \citet{WeissKim2024}.\newline\mbox{}}
\label{tab_sar1_ao}

\smallskip
\resizebox{.9\linewidth}{!}{
\begin{tabular}{lc@{\qquad}c@{\qquad}c@{\qquad}c}
\toprule
Delay & \multicolumn{2}{c}{``gilbert''} & \multicolumn{2}{c}{``Rong''} \\
$d$ & $\widehat{H}$ & $\hat{\uptau}$ & $\widehat{H}$ & $\hat{\uptau}$ \\
\midrule
2D-4.1 & \multicolumn{4}{l}{$(10, 10)$,\quad $\widetilde{\tau}$: {\it 0.151},\quad $\hat{\rho}(\1)$: {\it 0.029}} \\
\midrule
1 & 0.067 & 0.110 & 0.066 & 0.109 \\
2 & 0.053 & 0.066 & 0.052 & 0.066 \\
3 & 0.056 & 0.052 & 0.056 & 0.055 \\
4 & 0.055 & 0.049 & 0.056 & 0.049 \\
%5 & 0.056 & 0.060 & 0.055 & 0.060 \\
\midrule
2D-4.1 & \multicolumn{4}{l}{$(15, 15)$,\quad $\widetilde{\tau}$: {\it 0.271},\quad $\hat{\rho}(\1)$: {\it 0.048}} \\
\midrule
1 & 0.084 & 0.173 & 0.084 & 0.173 \\
2 & 0.052 & 0.047 & 0.052 & 0.048 \\
3 & 0.056 & 0.065 & 0.055 & 0.066 \\
4 & 0.055 & 0.055 & 0.053 & 0.055 \\
%5 & 0.054 & 0.067 & 0.054 & 0.067 \\
\midrule
2D-4.1 & \multicolumn{4}{l}{$(20, 20)$,\quad $\widetilde{\tau}$: {\it 0.496},\quad $\hat{\rho}(\1)$: {\it 0.050}} \\
\midrule
1 & 0.116 & 0.261 & 0.114 & 0.256 \\
2 & 0.053 & 0.059 & 0.052 & 0.060 \\
3 & 0.057 & 0.066 & 0.055 & 0.066 \\
4 & 0.053 & 0.056 & 0.053 & 0.057 \\
%5 & 0.053 & 0.059 & 0.053 & 0.062 \\
\midrule
2D-4.1 & \multicolumn{4}{l}{$(40, 25)$,\quad $\widetilde{\tau}$: {\it 0.872},\quad $\hat{\rho}(\1)$: {\it 0.087}} \\
\midrule
1 & 0.240 & 0.515 & 0.237 & 0.509 \\
2 & 0.052 & 0.063 & 0.052 & 0.066 \\
3 & 0.067 & 0.104 & 0.065 & 0.100 \\
4 & 0.052 & 0.057 & 0.054 & 0.063 \\
%5 & 0.052 & 0.061 & 0.056 & 0.073 \\
\midrule
2D-4.2 & \multicolumn{4}{l}{$(10, 10)$,\quad $\widetilde{\tau}$: {\it 0.142},\quad $\hat{\rho}(\1)$: {\it 0.031}} \\
\midrule
1 & 0.066 & 0.108 & 0.067 & 0.107 \\
2 & 0.053 & 0.065 & 0.054 & 0.065 \\
3 & 0.057 & 0.052 & 0.055 & 0.052 \\
4 & 0.057 & 0.048 & 0.054 & 0.049 \\
%5 & 0.055 & 0.059 & 0.057 & 0.060 \\
\midrule
2D-4.2 & \multicolumn{4}{l}{$(15, 15)$,\quad $\widetilde{\tau}$: {\it 0.274},\quad $\hat{\rho}(\1)$: {\it 0.043}} \\
\midrule
1 & 0.083 & 0.173 & 0.084 & 0.171 \\
2 & 0.052 & 0.047 & 0.052 & 0.048 \\
3 & 0.055 & 0.064 & 0.055 & 0.064 \\
4 & 0.054 & 0.056 & 0.052 & 0.056 \\
%5 & 0.053 & 0.066 & 0.055 & 0.067 \\
\midrule
2D-4.2 & \multicolumn{4}{l}{$(20, 20)$,\quad $\widetilde{\tau}$: {\it 0.491},\quad $\hat{\rho}(\1)$: {\it 0.054}} \\
\midrule
1 & 0.115 & 0.257 & 0.115 & 0.251 \\
2 & 0.051 & 0.060 & 0.051 & 0.060 \\
3 & 0.057 & 0.067 & 0.055 & 0.064 \\
4 & 0.053 & 0.057 & 0.051 & 0.057 \\
%5 & 0.053 & 0.059 & 0.053 & 0.062 \\
\midrule
2D-4.2 & \multicolumn{4}{l}{$(40, 25)$,\quad $\widetilde{\tau}$: {\it 0.874},\quad $\hat{\rho}(\1)$: {\it 0.098}} \\
\midrule
1 & 0.237 & 0.509 & 0.231 & 0.500 \\
2 & 0.052 & 0.063 & 0.052 & 0.063 \\
3 & 0.067 & 0.104 & 0.065 & 0.099 \\
4 & 0.053 & 0.057 & 0.053 & 0.063 \\
%5 & 0.053 & 0.062 & 0.056 & 0.074 \\
\bottomrule
\end{tabular}}
\end{multicols}
\end{table}

\begin{table}[ht!]
\centering
\begin{multicols}{2}
\caption{Simulated power for SQMA$(1,1)$-DGPs with sample size $(n_1,n_2)$, where Hilbert curve by \href{https://github.com/jakubcerveny/gilbert}{\nolinkurl{gilbert}} or \citet{rong21}, respectively. Italic numbers taken from \citet{WeissKim2024}.}
\label{tab_sqma11}

\smallskip
\resizebox{.9\linewidth}{!}{
\begin{tabular}{lc@{\qquad}c@{\qquad}c@{\qquad}c}
\toprule
Delay & \multicolumn{2}{c}{``gilbert''} & \multicolumn{2}{c}{``Rong''} \\
$d$ & $\widehat{H}$ & $\hat{\uptau}$ & $\widehat{H}$ & $\hat{\uptau}$ \\
\midrule
2D-5.1 & \multicolumn{4}{l}{$(10, 10)$,\quad $\widetilde{\tau}$: {\it 0.249},\quad $\hat{\rho}(\1)$: {\it 0.069}} \\
\midrule
1 & 0.105 & 0.236 & 0.096 & 0.209 \\
2 & 0.055 & 0.075 & 0.058 & 0.084 \\
3 & 0.060 & 0.062 & 0.061 & 0.068 \\
4 & 0.060 & 0.054 & 0.059 & 0.056 \\
%5 & 0.060 & 0.067 & 0.060 & 0.067 \\
\midrule
2D-5.1 & \multicolumn{4}{l}{$(15, 15)$,\quad $\widetilde{\tau}$: {\it 0.476},\quad $\hat{\rho}(\1)$: {\it 0.088}} \\
\midrule
1 & 0.149 & 0.353 & 0.152 & 0.356 \\
2 & 0.066 & 0.089 & 0.066 & 0.088 \\
3 & 0.063 & 0.085 & 0.064 & 0.086 \\
4 & 0.057 & 0.061 & 0.058 & 0.061 \\
%5 & 0.060 & 0.080 & 0.059 & 0.080 \\
\midrule
2D-5.1 & \multicolumn{4}{l}{$(20, 20)$,\quad $\widetilde{\tau}$: {\it 0.769},\quad $\hat{\rho}(\1)$: {\it 0.097}} \\
\midrule
1 & 0.324 & 0.638 & 0.279 & 0.577 \\
2 & 0.058 & 0.083 & 0.068 & 0.115 \\
3 & 0.067 & 0.098 & 0.066 & 0.094 \\
4 & 0.056 & 0.069 & 0.056 & 0.064 \\
%5 & 0.056 & 0.066 & 0.060 & 0.078 \\
\midrule
2D-5.1 & \multicolumn{4}{l}{$(40, 25)$,\quad $\widetilde{\tau}$: {\it 0.989},\quad $\hat{\rho}(\1)$: {\it 0.112}} \\
\midrule
1 & 0.696 & 0.927 & 0.610 & 0.879 \\
2 & 0.079 & 0.152 & 0.111 & 0.247 \\
3 & 0.103 & 0.207 & 0.098 & 0.187 \\
4 & 0.059 & 0.071 & 0.059 & 0.071 \\
%5 & 0.057 & 0.068 & 0.067 & 0.105 \\
\midrule
2D-5.2 & \multicolumn{4}{l}{$(10, 10)$,\quad $\widetilde{\tau}$: {\it 0.589},\quad $\hat{\rho}(\1)$: {\it 0.082}} \\
\midrule
1 & 0.143 & 0.345 & 0.128 & 0.305 \\
2 & 0.055 & 0.076 & 0.055 & 0.071 \\
3 & 0.073 & 0.099 & 0.065 & 0.081 \\
4 & 0.059 & 0.054 & 0.059 & 0.051 \\
%5 & 0.062 & 0.069 & 0.062 & 0.071 \\
\midrule
2D-5.2 & \multicolumn{4}{l}{$(15, 15)$,\quad $\widetilde{\tau}$: {\it 0.919},\quad $\hat{\rho}(\1)$: {\it 0.103}} \\
\midrule
1 & 0.323 & 0.657 & 0.307 & 0.637 \\
2 & 0.054 & 0.049 & 0.055 & 0.051 \\
3 & 0.068 & 0.106 & 0.071 & 0.109 \\
4 & 0.056 & 0.054 & 0.056 & 0.055 \\
%5 & 0.064 & 0.091 & 0.065 & 0.089 \\
\midrule
2D-5.2 & \multicolumn{4}{l}{$(20, 20)$,\quad $\widetilde{\tau}$: {\it 0.997},\quad $\hat{\rho}(\1)$: {\it 0.118}} \\
\midrule
1 & 0.582 & 0.876 & 0.519 & 0.830 \\
2 & 0.055 & 0.066 & 0.052 & 0.061 \\
3 & 0.098 & 0.183 & 0.080 & 0.135 \\
4 & 0.055 & 0.062 & 0.054 & 0.058 \\
%5 & 0.059 & 0.071 & 0.068 & 0.102 \\
\midrule
2D-5.2 & \multicolumn{4}{l}{$(40, 25)$,\quad $\widetilde{\tau}$: {\it 1.000},\quad $\hat{\rho}(\1)$: {\it 0.128}} \\
\midrule
1 & 0.996 & 1.000 & 0.941 & 0.995 \\
2 & 0.056 & 0.066 & 0.054 & 0.063 \\
3 & 0.087 & 0.155 & 0.122 & 0.253 \\
4 & 0.056 & 0.059 & 0.055 & 0.054 \\
%5 & 0.059 & 0.072 & 0.078 & 0.134 \\
\bottomrule
\end{tabular}}
%\end{table}

%\vfill\null
%\columnbreak

%\begin{table}{ht}
%\centering
\caption{Simulated power for multilateral SQMA$(1)$-DGPs with sample size $(n_1,n_2)$, where Hilbert curve by \href{https://github.com/jakubcerveny/gilbert}{\nolinkurl{gilbert}} or \citet{rong21}, respectively. Italic numbers taken from \citet{WeissKim2024}.\newline\mbox{}}
\label{tab_bsqma11}

\smallskip
\resizebox{.9\linewidth}{!}{
\begin{tabular}{lc@{\qquad}c@{\qquad}c@{\qquad}c}
\toprule
Delay & \multicolumn{2}{c}{``gilbert''} & \multicolumn{2}{c}{``Rong''} \\
$d$ & $\widehat{H}$ & $\hat{\uptau}$ & $\widehat{H}$ & $\hat{\uptau}$ \\
\midrule
2D-6.1 & \multicolumn{4}{l}{$(10, 10)$,\quad $\widetilde{\tau}$: {\it 0.213},\quad $\hat{\rho}(\1)$: {\it 0.169}} \\
\midrule
1 & 0.084 & 0.146 & 0.077 & 0.131 \\
2 & 0.053 & 0.058 & 0.057 & 0.082 \\
3 & 0.088 & 0.158 & 0.085 & 0.147 \\
4 & 0.067 & 0.100 & 0.078 & 0.134 \\
%5 & 0.071 & 0.109 & 0.065 & 0.087 \\
\midrule
2D-6.1 & \multicolumn{4}{l}{$(15, 15)$,\quad  $\widetilde{\tau}$: {\it 0.203},\quad $\hat{\rho}(\1)$: {\it 0.203}} \\
\midrule
1 & 0.085 & 0.165 & 0.085 & 0.163 \\
2 & 0.100 & 0.188 & 0.099 & 0.187 \\
3 & 0.086 & 0.167 & 0.086 & 0.169 \\
4 & 0.136 & 0.304 & 0.136 & 0.303 \\
%5 & 0.073 & 0.129 & 0.073 & 0.128 \\
\midrule
2D-6.1 & \multicolumn{4}{l}{$(20, 20)$,\quad $\widetilde{\tau}$: {\it 0.235},\quad $\hat{\rho}(\1)$: {\it 0.229}} \\
\midrule
1 & 0.167 & 0.362 & 0.138 & 0.302 \\
2 & 0.049 & 0.051 & 0.066 & 0.107 \\
3 & 0.166 & 0.336 & 0.133 & 0.270 \\
4 & 0.106 & 0.232 & 0.121 & 0.267 \\
%5 & 0.097 & 0.187 & 0.080 & 0.141 \\
\midrule
2D-6.1 & \multicolumn{4}{l}{$(40, 25)$,\quad $\widetilde{\tau}$: {\it 0.258},\quad $\hat{\rho}(\1)$: {\it 0.238}} \\
\midrule
1 & 0.290 & 0.536 & 0.194 & 0.389 \\
2 & 0.076 & 0.150 & 0.281 & 0.571 \\
3 & 0.347 & 0.629 & 0.225 & 0.454 \\
4 & 0.339 & 0.646 & 0.503 & 0.801 \\
%5 & 0.140 & 0.306 & 0.104 & 0.217 \\
\midrule
2D-6.2 & \multicolumn{4}{l}{$(10, 10)$,\quad $\widetilde{\tau}$: {\it 0.391},\quad $\hat{\rho}(\1)$: {\it 0.147}} \\
\midrule
1 & 0.076 & 0.134 & 0.083 & 0.149 \\
2 & 0.055 & 0.072 & 0.059 & 0.093 \\
3 & 0.061 & 0.081 & 0.059 & 0.068 \\
4 & 0.057 & 0.066 & 0.056 & 0.055 \\
%5 & 0.061 & 0.074 & 0.057 & 0.066 \\
\midrule
2D-6.2 & \multicolumn{4}{l}{$(15, 15)$,\quad $\widetilde{\tau}$: {\it 0.660},\quad $\hat{\rho}(\1)$: {\it 0.179}} \\
\midrule
1 & 0.133 & 0.285 & 0.135 & 0.285 \\
2 & 0.081 & 0.146 & 0.084 & 0.148 \\
3 & 0.063 & 0.097 & 0.063 & 0.098 \\
4 & 0.056 & 0.068 & 0.057 & 0.068 \\
%5 & 0.057 & 0.078 & 0.057 & 0.077 \\
\midrule
2D-6.2 & \multicolumn{4}{l}{$(20, 20)$,\quad $\widetilde{\tau}$: {\it 0.888},\quad $\hat{\rho}(\1)$: {\it 0.201}} \\
\midrule
1 & 0.148 & 0.333 & 0.181 & 0.400 \\
2 & 0.056 & 0.079 & 0.087 & 0.187 \\
3 & 0.076 & 0.124 & 0.067 & 0.107 \\
4 & 0.072 & 0.123 & 0.055 & 0.066 \\
%5 & 0.067 & 0.099 & 0.060 & 0.082 \\
\midrule
2D-6.2 & \multicolumn{4}{l}{$(40, 25)$,\quad $\widetilde{\tau}$: {\it 0.998},\quad $\hat{\rho}(\1)$: {\it 0.214}} \\
\midrule
1 & 0.382 & 0.665 & 0.486 & 0.766 \\
2 & 0.105 & 0.233 & 0.226 & 0.495 \\
3 & 0.126 & 0.273 & 0.088 & 0.173 \\
4 & 0.068 & 0.109 & 0.056 & 0.070 \\
%5 & 0.072 & 0.126 & 0.063 & 0.098 \\
\bottomrule
\end{tabular}}
\end{multicols}
\end{table}

\begin{table}[ht!]
\centering
\begin{multicols}{2}
\caption{Simulated power for SQMA$(1,1,1)$-DGPs with sample size $(n_1,n_2,n_3)$, where Hilbert curve by \href{https://github.com/jakubcerveny/gilbert}{\nolinkurl{gilbert}} or \citet{rong21}, respectively.}
\label{tab_sqma111}

\smallskip
\resizebox{.9\linewidth}{!}{
\begin{tabular}{lc@{\qquad}c@{\qquad}c@{\qquad}c}
\toprule
Delay & \multicolumn{2}{c}{``gilbert''} & \multicolumn{2}{c}{``Rong''} \\
$d$ & $\widehat{H}$ & $\hat{\uptau}$ & $\widehat{H}$ & $\hat{\uptau}$ \\
\midrule
3D-1.1 & \multicolumn{4}{l}{$(7, 7, 7)$,\quad $\hat{\rho}(\1)$: 0.099} \\
\midrule
1 & 0.595 & 0.879 & 0.666 & 0.915 \\
2 & 0.059 & 0.093 & 0.056 & 0.077 \\
3 & 0.156 & 0.325 & 0.138 & 0.285 \\
4 & 0.144 & 0.298 & 0.148 & 0.304 \\
%5 & 0.102 & 0.208 & 0.149 & 0.329 \\
\midrule
3D-1.1 & \multicolumn{4}{l}{$(10, 10, 7)$,\quad $\hat{\rho}(\1)$: 0.120} \\
\midrule
1 & 0.899 & 0.989 & 1.000 & 1.000 \\
2 & 0.074 & 0.138 & 0.054 & 0.091 \\
3 & 0.211 & 0.433 & 0.059 & 0.097 \\
4 & 0.184 & 0.388 & 0.061 & 0.095 \\
%5 & 0.159 & 0.346 & 0.059 & 0.093 \\
\midrule
3D-1.1 & \multicolumn{4}{l}{$(10, 15, 7)$,\quad $\hat{\rho}(\1)$: 0.132} \\
\midrule
1 & 0.982 & 0.999 & 0.992 & 1.000 \\
2 & 0.069 & 0.125 & 0.059 & 0.092 \\
3 & 0.342 & 0.625 & 0.340 & 0.616 \\
4 & 0.316 & 0.592 & 0.294 & 0.562 \\
%5 & 0.278 & 0.551 & 0.294 & 0.576 \\
\midrule
3D-1.1 & \multicolumn{4}{l}{$(15, 15, 7)$,\quad $\hat{\rho}(\1)$: 0.142} \\
\midrule
1 & 0.999 & 1.000 & 1.000 & 1.000 \\
2 & 0.075 & 0.142 & 0.055 & 0.075 \\
3 & 0.554 & 0.825 & 0.465 & 0.745 \\
4 & 0.488 & 0.770 & 0.495 & 0.774 \\
%5 & 0.282 & 0.548 & 0.524 & 0.805 \\
\midrule
3D-1.2 & \multicolumn{4}{l}{$(7, 7, 7)$,\quad $\hat{\rho}(\1)$: 0.056} \\
\midrule
1 & 0.318 & 0.574 & 0.276 & 0.513 \\
2 & 0.064 & 0.104 & 0.055 & 0.074 \\
3 & 0.092 & 0.177 & 0.100 & 0.201 \\
4 & 0.228 & 0.478 & 0.229 & 0.478 \\
%5 & 0.063 & 0.081 & 0.057 & 0.066 \\
\midrule
3D-1.2 & \multicolumn{4}{l}{$(10, 10, 7)$,\quad $\hat{\rho}(\1)$: 0.070} \\
\midrule
1 & 0.477 & 0.750 & 0.077 & 0.106 \\
2 & 0.088 & 0.180 & 0.056 & 0.068 \\
3 & 0.090 & 0.182 & 0.073 & 0.132 \\
4 & 0.292 & 0.576 & 0.054 & 0.070 \\
%5 & 0.055 & 0.063 & 0.071 & 0.120 \\
\midrule
3D-1.2 & \multicolumn{4}{l}{$(10, 15, 7)$,\quad $\hat{\rho}(\1)$: 0.077} \\
\midrule
1 & 0.716 & 0.910 & 0.669 & 0.882 \\
2 & 0.085 & 0.172 & 0.061 & 0.097 \\
3 & 0.159 & 0.332 & 0.182 & 0.376 \\
4 & 0.554 & 0.833 & 0.522 & 0.810 \\
%5 & 0.059 & 0.079 & 0.058 & 0.073 \\
\midrule
3D-1.2 & \multicolumn{4}{l}{$(15, 15, 7)$,\quad $\hat{\rho}(\1)$: 0.084} \\
\midrule
1 & 0.889 & 0.979 & 0.825 & 0.958 \\
2 & 0.093 & 0.193 & 0.056 & 0.075 \\
3 & 0.220 & 0.443 & 0.264 & 0.515 \\
4 & 0.803 & 0.961 & 0.807 & 0.961 \\
%5 & 0.080 & 0.143 & 0.058 & 0.068 \\
\bottomrule
\end{tabular}}
%\end{table}

%\vfill\null
%\columnbreak

%\begin{table}{ht}
%\centering
\caption{Simulated power for multilateral SQMA$(1)$-DGPs with sample size $(n_1,n_2,n_3)$, where Hilbert curve by \href{https://github.com/jakubcerveny/gilbert}{\nolinkurl{gilbert}} or \citet{rong21}, respectively.}
\label{tab_bsqma111}

\smallskip
\resizebox{.9\linewidth}{!}{
\begin{tabular}{lc@{\qquad}c@{\qquad}c@{\qquad}c}
\toprule
Delay & \multicolumn{2}{c}{``gilbert''} & \multicolumn{2}{c}{``Rong''} \\
$d$ & $\widehat{H}$ & $\hat{\uptau}$ & $\widehat{H}$ & $\hat{\uptau}$ \\
\midrule
3D-2.1 & \multicolumn{4}{l}{$(7, 7, 7)$,\quad $\hat{\rho}(\1)$: 0.149} \\
\midrule
1 & 0.069 & 0.096 & 0.082 & 0.128 \\
2 & 0.059 & 0.081 & 0.063 & 0.099 \\
3 & 0.071 & 0.107 & 0.064 & 0.080 \\
4 & 0.293 & 0.567 & 0.304 & 0.576 \\
%5 & 0.085 & 0.138 & 0.262 & 0.484 \\
\midrule
3D-2.1 & \multicolumn{4}{l}{$(10, 10, 7)$,\quad $\hat{\rho}(\1)$: 0.182} \\
\midrule
1 & 0.072 & 0.119 & 0.989 & 0.999 \\
2 & 0.074 & 0.125 & 1.000 & 1.000 \\
3 & 0.071 & 0.117 & 0.091 & 0.110 \\
4 & 0.115 & 0.250 & 0.081 & 0.100 \\
%5 & 0.079 & 0.133 & 0.094 & 0.128 \\
\midrule
3D-2.1 & \multicolumn{4}{l}{$(10, 15, 7)$,\quad $\hat{\rho}(\1)$: 0.198} \\
\midrule
1 & 0.071 & 0.104 & 0.100 & 0.169 \\
2 & 0.076 & 0.142 & 0.057 & 0.084 \\
3 & 0.082 & 0.134 & 0.068 & 0.089 \\
4 & 0.417 & 0.705 & 0.609 & 0.859 \\
%5 & 0.125 & 0.255 & 0.419 & 0.701 \\
\midrule
3D-2.1 & \multicolumn{4}{l}{$(15, 15, 7)$,\quad $\hat{\rho}(\1)$: 0.211} \\
\midrule
1 & 0.075 & 0.114 & 0.127 & 0.235 \\
2 & 0.069 & 0.105 & 0.080 & 0.148 \\
3 & 0.098 & 0.174 & 0.069 & 0.083 \\
4 & 0.897 & 0.984 & 0.910 & 0.986 \\
%5 & 0.179 & 0.368 & 0.813 & 0.958 \\
\midrule
3D-2.2 & \multicolumn{4}{l}{$(7, 7, 7)$,\quad $\hat{\rho}(\1)$: 0.084} \\
\midrule
1 & 0.056 & 0.060 & 0.058 & 0.072 \\
2 & 0.054 & 0.065 & 0.057 & 0.071 \\
3 & 0.055 & 0.062 & 0.057 & 0.067 \\
4 & 0.106 & 0.222 & 0.111 & 0.235 \\
%5 & 0.054 & 0.059 & 0.079 & 0.138 \\
\midrule
3D-2.2 & \multicolumn{4}{l}{$(10, 10, 7)$,\quad $\hat{\rho}(\1)$: 0.107} \\
\midrule
1 & 0.057 & 0.069 & 0.197 & 0.381 \\
2 & 0.060 & 0.083 & 0.219 & 0.447 \\
3 & 0.053 & 0.057 & 0.065 & 0.076 \\
4 & 0.073 & 0.136 & 0.062 & 0.073 \\
%5 & 0.052 & 0.059 & 0.065 & 0.078 \\
\midrule
3D-2.2 & \multicolumn{4}{l}{$(10, 15, 7)$,\quad $\hat{\rho}(\1)$: 0.118} \\
\midrule
1 & 0.056 & 0.064 & 0.061 & 0.076 \\
2 & 0.057 & 0.082 & 0.053 & 0.067 \\
3 & 0.054 & 0.059 & 0.058 & 0.070 \\
4 & 0.164 & 0.352 & 0.279 & 0.547 \\
%5 & 0.056 & 0.075 & 0.062 & 0.092 \\
\midrule
3D-2.2 & \multicolumn{4}{l}{$(15, 15, 7)$,\quad $\hat{\rho}(\1)$: 0.126} \\
\midrule
1 & 0.056 & 0.064 & 0.068 & 0.098 \\
2 & 0.057 & 0.072 & 0.059 & 0.082 \\
3 & 0.054 & 0.062 & 0.063 & 0.088 \\
4 & 0.379 & 0.674 & 0.446 & 0.740 \\
%5 & 0.053 & 0.066 & 0.118 & 0.256 \\
\bottomrule
\end{tabular}}
\end{multicols}
\end{table}

\begin{table}[ht!]
\centering
\begin{multicols}{2}
\caption{Simulated power for GRF-DGPs with sample size $(n_1,n_2,n_3)$, where Hilbert curve by \href{https://github.com/jakubcerveny/gilbert}{\nolinkurl{gilbert}} or \citet{rong21}, respectively.}
\label{tab_grf}

\smallskip
\resizebox{.9\linewidth}{!}{
\begin{tabular}{lc@{\qquad}c@{\qquad}c@{\qquad}c}
\toprule
Delay & \multicolumn{2}{c}{``gilbert''} & \multicolumn{2}{c}{``Rong''} \\
$d$ & $\widehat{H}$ & $\hat{\uptau}$ & $\widehat{H}$ & $\hat{\uptau}$ \\
\midrule
3D-3.1 & \multicolumn{4}{l}{$(7, 7, 7)$,\quad $\hat{\rho}(\1)$: 0.049} \\
\midrule
1 & 0.125 & 0.285 & 0.115 & 0.257 \\
2 & 0.051 & 0.058 & 0.058 & 0.074 \\
3 & 0.066 & 0.099 & 0.077 & 0.136 \\
4 & 0.056 & 0.071 & 0.062 & 0.093 \\
%5 & 0.058 & 0.078 & 0.060 & 0.096 \\
\midrule
3D-3.1 & \multicolumn{4}{l}{$(10, 10, 7)$,\quad $\hat{\rho}(\1)$: 0.092} \\
\midrule
1 & 0.221 & 0.485 & 0.616 & 0.882 \\
2 & 0.053 & 0.061 & 0.071 & 0.129 \\
3 & 0.062 & 0.089 & 0.056 & 0.065 \\
4 & 0.056 & 0.069 & 0.053 & 0.062 \\
%5 & 0.069 & 0.122 & 0.055 & 0.064 \\
\midrule
3D-3.1 & \multicolumn{4}{l}{$(10, 15, 7)$,\quad $\hat{\rho}(\1)$: 0.131} \\
\midrule
1 & 0.310 & 0.618 & 0.356 & 0.673 \\
2 & 0.051 & 0.061 & 0.058 & 0.081 \\
3 & 0.087 & 0.165 & 0.070 & 0.115 \\
4 & 0.065 & 0.105 & 0.054 & 0.062 \\
%5 & 0.084 & 0.170 & 0.066 & 0.110 \\
\midrule
3D-3.1 & \multicolumn{4}{l}{$(15, 15, 7)$,\quad $\hat{\rho}(\1)$: 0.190} \\
\midrule
1 & 0.470 & 0.780 & 0.541 & 0.833 \\
2 & 0.052 & 0.061 & 0.066 & 0.107 \\
3 & 0.118 & 0.247 & 0.076 & 0.133 \\
4 & 0.064 & 0.101 & 0.054 & 0.060 \\
%5 & 0.071 & 0.125 & 0.074 & 0.134 \\
\midrule
3D-3.2 & \multicolumn{4}{l}{$(7, 7, 7)$,\quad $\hat{\rho}(\1)$: 0.520} \\
\midrule
1 & 0.379 & 0.703 & 0.380 & 0.708 \\
2 & 0.056 & 0.078 & 0.075 & 0.127 \\
3 & 0.128 & 0.264 & 0.179 & 0.384 \\
4 & 0.095 & 0.180 & 0.147 & 0.303 \\
%5 & 0.101 & 0.211 & 0.156 & 0.335 \\
\midrule
3D-3.2 & \multicolumn{4}{l}{$(10, 10, 7)$,\quad $\hat{\rho}(\1)$: 0.837} \\
\midrule
1 & 0.703 & 0.928 & 0.996 & 1.000 \\
2 & 0.066 & 0.110 & 0.446 & 0.735 \\
3 & 0.111 & 0.227 & 0.136 & 0.266 \\
4 & 0.095 & 0.180 & 0.083 & 0.143 \\
%5 & 0.201 & 0.430 & 0.085 & 0.150 \\
\midrule
3D-3.2 & \multicolumn{4}{l}{$(10, 15, 7)$,\quad $\hat{\rho}(\1)$: 0.949} \\
\midrule
1 & 0.874 & 0.983 & 0.911 & 0.991 \\
2 & 0.056 & 0.077 & 0.098 & 0.209 \\
3 & 0.255 & 0.515 & 0.166 & 0.356 \\
4 & 0.204 & 0.415 & 0.086 & 0.155 \\
%5 & 0.339 & 0.631 & 0.180 & 0.384 \\
\midrule
3D-3.2 & \multicolumn{4}{l}{$(15, 15, 7)$,\quad $\hat{\rho}(\1)$: 0.992} \\
\midrule
1 & 0.974 & 0.999 & 0.988 & 1.000 \\
2 & 0.066 & 0.109 & 0.156 & 0.342 \\
3 & 0.458 & 0.751 & 0.215 & 0.438 \\
4 & 0.222 & 0.446 & 0.098 & 0.185 \\
%5 & 0.264 & 0.516 & 0.252 & 0.493 \\
\bottomrule
\end{tabular}}
%\end{table}

%\vfill\null
%\columnbreak

%\begin{table}{ht}
%\centering
\caption{Simulated power for GRF-DGPs with sample size $(n_1,n_2,n_3)$ like in Table~\ref{tab_grf}, but data contaminated by AOs.\newline\mbox{}}
\label{tab_grf_ao}

\smallskip
\resizebox{.9\linewidth}{!}{
\begin{tabular}{lc@{\qquad}c@{\qquad}c@{\qquad}c}
\toprule
Delay & \multicolumn{2}{c}{``gilbert''} & \multicolumn{2}{c}{``Rong''} \\
$d$ & $\widehat{H}$ & $\hat{\uptau}$ & $\widehat{H}$ & $\hat{\uptau}$ \\
\midrule
3D-4.1 & \multicolumn{4}{l}{$(7, 7, 7)$,\quad $\hat{\rho}(\1)$: 0.016} \\
\midrule
1 & 0.085 & 0.176 & 0.083 & 0.163 \\
2 & 0.051 & 0.056 & 0.054 & 0.063 \\
3 & 0.058 & 0.075 & 0.065 & 0.097 \\
4 & 0.055 & 0.062 & 0.057 & 0.075 \\
%5 & 0.054 & 0.069 & 0.057 & 0.077 \\
\midrule
3D-4.1 & \multicolumn{4}{l}{$(10, 10, 7)$,\quad $\hat{\rho}(\1)$: 0.021} \\
\midrule
1 & 0.127 & 0.288 & 0.336 & 0.644 \\
2 & 0.053 & 0.057 & 0.061 & 0.095 \\
3 & 0.056 & 0.074 & 0.053 & 0.059 \\
4 & 0.053 & 0.061 & 0.051 & 0.056 \\
%5 & 0.060 & 0.089 & 0.053 & 0.058 \\
\midrule
3D-4.1 & \multicolumn{4}{l}{$(10, 15, 7)$,\quad $\hat{\rho}(\1)$: 0.025} \\
\midrule
1 & 0.170 & 0.381 & 0.190 & 0.425 \\
2 & 0.051 & 0.059 & 0.055 & 0.067 \\
3 & 0.068 & 0.110 & 0.061 & 0.085 \\
4 & 0.056 & 0.078 & 0.052 & 0.056 \\
%5 & 0.067 & 0.113 & 0.060 & 0.084 \\
\midrule
3D-4.1 & \multicolumn{4}{l}{$(15, 15, 7)$,\quad $\hat{\rho}(\1)$: 0.027} \\
\midrule
1 & 0.243 & 0.514 & 0.286 & 0.576 \\
2 & 0.051 & 0.055 & 0.057 & 0.081 \\
3 & 0.081 & 0.152 & 0.062 & 0.093 \\
4 & 0.059 & 0.076 & 0.052 & 0.054 \\
%5 & 0.061 & 0.092 & 0.061 & 0.094 \\
\midrule
3D-4.2 & \multicolumn{4}{l}{$(7, 7, 7)$,\quad $\hat{\rho}(\1)$: 0.023} \\
\midrule
1 & 0.202 & 0.446 & 0.200 & 0.448 \\
2 & 0.053 & 0.068 & 0.063 & 0.095 \\
3 & 0.088 & 0.164 & 0.112 & 0.233 \\
4 & 0.073 & 0.121 & 0.098 & 0.195 \\
%5 & 0.076 & 0.140 & 0.101 & 0.211 \\
\midrule
3D-4.2 & \multicolumn{4}{l}{$(10, 10, 7)$,\quad $\hat{\rho}(\1)$: 0.035} \\
\midrule
1 & 0.393 & 0.715 & 0.898 & 0.986 \\
2 & 0.057 & 0.085 & 0.247 & 0.508 \\
3 & 0.079 & 0.147 & 0.091 & 0.171 \\
4 & 0.072 & 0.122 & 0.066 & 0.102 \\
%5 & 0.122 & 0.265 & 0.067 & 0.105 \\
\midrule
3D-4.2 & \multicolumn{4}{l}{$(10, 15, 7)$,\quad $\hat{\rho}(\1)$: 0.046} \\
\midrule
1 & 0.558 & 0.851 & 0.616 & 0.884 \\
2 & 0.052 & 0.067 & 0.073 & 0.136 \\
3 & 0.143 & 0.310 & 0.105 & 0.216 \\
4 & 0.122 & 0.258 & 0.067 & 0.106 \\
%5 & 0.187 & 0.405 & 0.112 & 0.239 \\
\midrule
3D-4.2 & \multicolumn{4}{l}{$(15, 15, 7)$,\quad $\hat{\rho}(\1)$: 0.064} \\
\midrule
1 & 0.762 & 0.950 & 0.824 & 0.969 \\
2 & 0.058 & 0.084 & 0.101 & 0.212 \\
3 & 0.242 & 0.494 & 0.126 & 0.267 \\
4 & 0.129 & 0.273 & 0.072 & 0.121 \\
%5 & 0.147 & 0.321 & 0.144 & 0.308 \\
\bottomrule
\end{tabular}}
\end{multicols}
\end{table}

\end{document}